\documentclass[]{interact}

\usepackage{epstopdf}
\usepackage[caption=false]{subfig}

\usepackage[numbers,sort&compress]{natbib}
\bibpunct[, ]{[}{]}{,}{n}{,}{,}

\theoremstyle{plain}
\usepackage{amssymb}
\usepackage{amsmath}

\usepackage{float}
\usepackage{amsthm}
\usepackage{booktabs}
\usepackage{multirow}
\usepackage{xurl}
\usepackage{mathtools}
\usepackage{adjustbox}
\usepackage{lmodern}
\usepackage{tabularx}
\usepackage[usenames,dvipsnames]{color}
\usepackage{bm}
\usepackage{bbm}
\usepackage[linesnumbered]{algorithm2e} 
\renewcommand{\P}{\mathbb{P}}
\newcommand{\E}{\mathbb{E}}

\newcommand{\argmin}{\operatornamewithlimits{argmin}}
\newcommand{\argmax}{\operatornamewithlimits{argmax}} 

\RestyleAlgo{ruled} 
\begin{document}

\setlength{\parindent}{15pt}

\title{Data-Adaptive Automatic Threshold Calibration for Stability Selection}

\author{\name{Martin Huang\textsuperscript{a}\thanks{Corresponding Author: Martin Huang. Email: martin.huang@sydney.edu.au}, Samuel Muller\textsuperscript{a,b}, Garth Tarr\textsuperscript{a}}
\affil{\textsuperscript{a}The University of Sydney, School of Mathematics and Statistics, Sydney, 2006, New South Wales, Australia;\\ \textsuperscript{b}Macquarie University, Faculty of Science and Engineering, Sydney, 2109, New South Wales, Australia}
}

\maketitle
\begin{abstract}
Stability selection has gained popularity as a method for enhancing the performance of variable selection algorithms while controlling false discovery rates. However, achieving these desirable properties depends on correctly specifying the stable threshold parameter, which can be challenging. An arbitrary choice of this parameter can substantially alter the set of selected variables, as the variables' selection probabilities are inherently data-dependent. To address this issue, we propose Exclusion Automatic Threshold Selection (EATS), a data-adaptive algorithm that streamlines stability selection by automating the threshold specification process. EATS initially filters out potential noise variables using an exclusion probability threshold, derived from applying stability selection to a randomly shuffled version of the dataset. Following this, EATS selects the stable threshold parameter using the elbow method, balancing the marginal utility of including additional variables against the risk of selecting superfluous variables. We evaluate our approach through an extensive simulation study, benchmarking across commonly used variable selection algorithms and static stable threshold values.
\end{abstract}

\section{Introduction}\label{chap:introduction}
\pagenumbering{arabic}

Difficulties in variable selection arise when the number of predictor variables far surpasses the number of samples, often described as high-dimensional data. High-dimensional datasets are not only incompatible for traditional linear models but may also contain low signal, causing variable selection algorithms to misidentify signal and noise variables. Stability selection addresses this issue by assigning selection probabilities to each variable \citep{meinshausen_stability_2010}. These selection probabilities are empirically estimated through the repeated fitting of a variable selection algorithm, assisting users in identifying and selecting a reliable set of estimated signal variables.

To determine whether or not a variable is stable, a threshold selection probability $\pi$ needs to be initially defined. Then, any variable with an empirical selection probability greater than $\pi$ is considered to be stable and estimated as a signal variable. These empirical selection probabilities are dependent on a multitude of factors. Such factors include the strength of the signal between the predictor variables in the design matrix $\bm{X}$ and the outcome $\bm{y}$, the dimensions and correlation structure of $\bm{X}$, and the subsample taken from the design matrix \citep{buhlmann_high-dimensional_2014, meinshausen_stability_2010}. As we see later in Section \ref{sec:simulation}, a dataset with a low signal-to-noise ratio (SNR) or small sample size may yield small selection probabilities for signal variables, which can lead to false negatives under a fixed threshold.

\begin{figure}
    \centering
    \includegraphics[scale = 0.3]{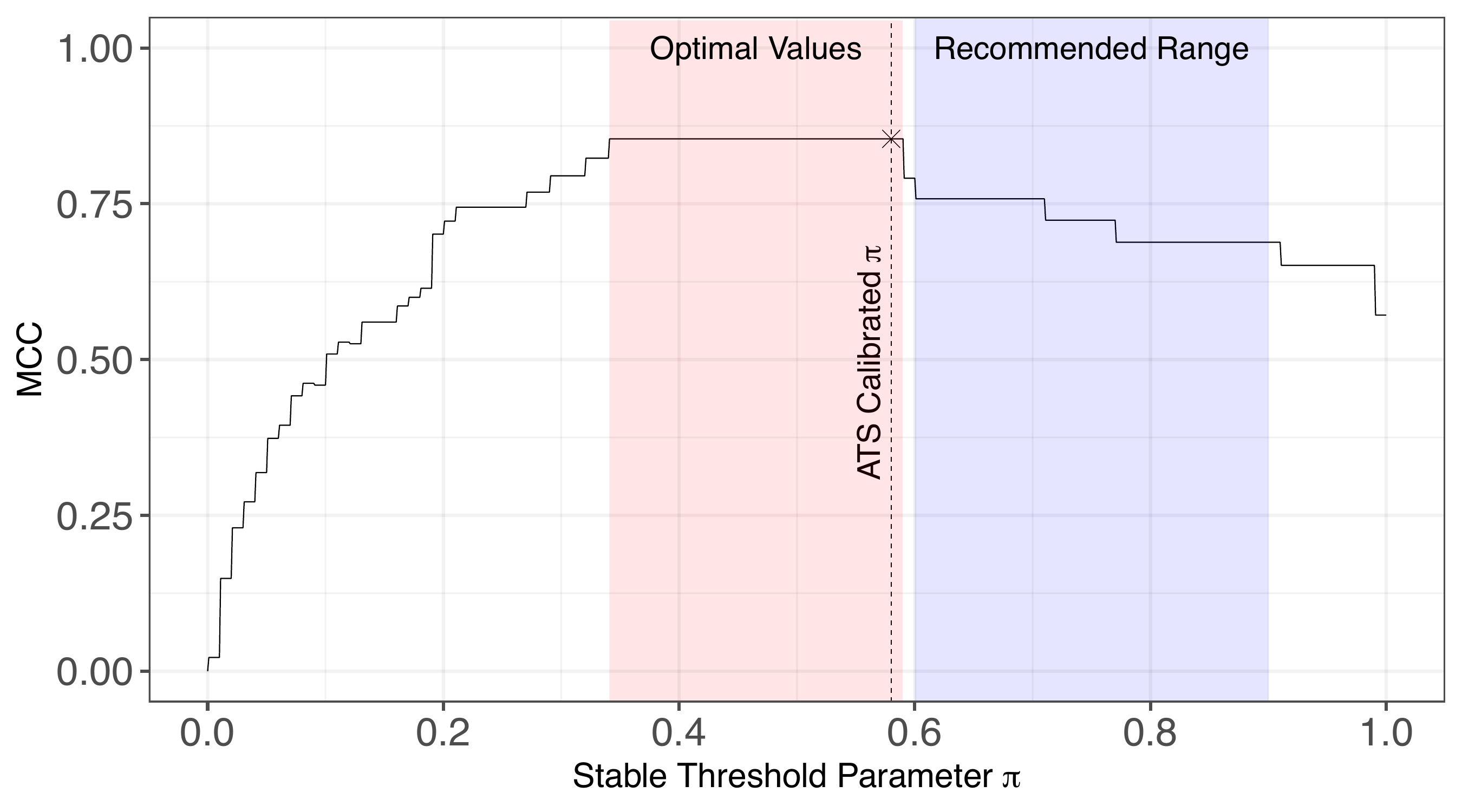}
    \caption{An example of the effect on signal variable recovery, represented by Matthew's correlation coefficient (MCC), for a change in the stable threshold parameter, $\pi$. Formal definitions of the different methods are introduced in Section \ref{chap:methodology}. In this setting, the recommended range for $\pi \in [0.6,0.9]$ from \citet{meinshausen_stability_2010} does not overlap with the range of $\pi$ that achieves the maximum MCC. Our proposed method EATS returns a calibrated value of $\hat{\pi}$ in the optimal MCC range without the need to manually specify the parameter.}
    \label{fig:piexample}
\end{figure}

Rather than specifying a fixed stability threshold parameter $\pi$, our method adapts to the empirical selection probabilities and is thus data-adaptive. A simple example of the effect of accuracy across a range of $\pi$ can be seen in Figure \ref{fig:piexample}. The goal is to estimate $\pi$ such that it maximises recovery of signal variables. However, this example demonstrates that the recommended range of $\pi \in [0.6, 0.9]$ from \citet{meinshausen_stability_2010} is not always optimal, as the $\pi$ values that maximise Matthew's correlation coefficient (MCC) actually fall outside this interval.

There have been several advances and alternatives to the original stability selection framework proposed by \citet{meinshausen_stability_2010}. Recently, \citet{bodinier_automated_2023} suggested an automated calibration procedure by maximising a stability score to concurrently estimate the penalised regression parameter $\lambda$, and identify a stable threshold parameter $\pi$. In \citet{hedou_discovery_2024}, the main ideology of stability selection is kept, however, artificial predictor variables are concatenated to the original dataset to inject noise for the discovery of sparse and reliable omic biomarkers. \citet{maddu_stability_2022} combined stability selection with the compressed sensing iterative hard-thresholding algorithm, enabling a fast and robust framework for equation inference in noisy spatiotemporal data. \citet{melikechi_2024} introduced an alternative approach to stability selection based on integrating stability paths rather than maximising over them, thereby providing much stronger upper bounds on the expected number of false positives. As a last example, \citet{gauraha_post_2017} provided a two-stage stability selection procedure that is consistent in variable selection under the generalised irrepresentable condition instead of the stronger sparse eigenvalue condition. 

In addition to these advancements in stability selection, many domains outside of statistics have benefited from its properties. For instance, \citet{lu_performance_2017} applied stability selection to investigate diverse chemical data to identify a subset of measured variables that can effectively distinguish between groups in metabolomics. \citet{alexander_stability_2011} utilised stability selection in genome-wide association studies to estimate indicators for common conditions such as bipolar disorder and rheumatoid arthritis.  In addition to signal variable recovery, \citet{mordelet_stability_2013} used stability selection to develop DNA-binding regression models.

Heuristic and visual processes for choosing parameters with respect to a metric are commonly found in the statistical machine learning literature, such as the investigative scree plots or regularisation paths \citep{cattell_scree_1966, tibshirani_regression_1996}. These methods aim to optimise a parameter based on the changing marginal benefit and metric trade-off. More specifically, the elbow method is frequently used to determine the optimal value of $k$ in $k$-means clustering or $k$-nearest neighbours, requiring the visualisation of the relationship between the parameter and an error metric, such as the sum of squares \citep{fix_discriminatory_1989, macqueen_methods_1967, thorndike_who_1953}. The elbow method is also commonly used in principal component analysis (PCA), where the parameter of interest is the number of dimensions, and the metric compared is the dimensions' corresponding eigenvalues.

 With the introduced elbow method, we propose EATS, a data-adaptive and automatic stable threshold estimation algorithm that utilises the marginal utility trade-off philosophy. To automate the elbow method, we extend the work of \citet{zhu_automatic_2006} and modify it for stability selection.

The remainder of this paper is structured as follows. Section \ref{chap:prelims} introduces notation and the statistical background required for EATS. Section \ref{chap:methodology} outlines EATS' framework and algorithmic implementation, including guidance to implement false discovery properties. Section \ref{chap:numerical} investigates the performance of stability selection for a variety of dimensions, signal-to-noise ratios, and stable threshold parameters, benchmarked against common variable selection algorithms. Finally, we revisit the important findings in Section \ref{chap:conclusion}.

\section{Preliminaries}\label{chap:prelims}

Let $\bm{Z} = (\bm{X}, \bm{y})$ denote a dataset, where $\bm{y} \in \mathbb{R}^n$ is the response $n$-vector and $\bm{X}\in \mathbb{R}^{n\times p}$ the design matrix. The $i$th row in $\bm{Z}$ is then denoted as $\bm{z}_i = (\bm{x}_i, y_i)$. Consider a linear regression framework $\bm{y} = \bm{X}\bm{\beta} + \bm{\epsilon}$, where $\bm{\epsilon}$ is a vector of random noise and $\bm{\beta}^\top = (\beta_1, \dots, \beta_p)$ is the unknown population coefficient vector. Without loss of generality, we centre the predictors and response variables such that an intercept is not required.  In $\bm{\beta}$ we define two subsets; $\bm{\beta}_\mathcal{S}$ the set of unknown coefficients for signal variables, and $\bm{\beta}_{\mathcal{N}}$, the coefficient vector for noise variables. By definition, $\bm{\beta}_{\mathcal{N}} = \bm{0}$ and elements in $\bm{\beta}_\mathcal{S}$ are non-zero. The goal of variable selection does not necessarily require consistent estimation of $\bm{\beta}$, i.e.\ non-zero regression coefficients only need to be estimated as non-zero whereas the zero regression coefficients need to be consistently estimated. Define the index set of non-zero values in $\bm{\beta}$ to be $\mathcal{S} = \{k: \beta_k \neq 0,\ k =1,\dots,p\}$ and the index set of noise variables with a zero coefficient by $\mathcal{N} = \{k :\beta_k = 0,\ k = 1,\dots,p\}$. We then define $\bm{X}_\mathcal{S} \subseteq \bm{X}$ as the subset of the original design matrix which contains all signal variables and $\bm{X}_{\mathcal{N}} \subseteq \bm{X}$ be the subset of the original design matrix containing only noise predictor variables.

A common variable selection algorithm used in stability selection to estimate $\mathcal{S}$ and $\mathcal{N}$ is the Least Absolute Shrinkage and Selection Operator (LASSO) \citep{tibshirani_regression_1996}, which is the solution to
$\bm{\hat{\beta}} ^\lambda = \argmin_{\bm{\beta}}\left(\lVert \bm{y} - \bm{X} \bm{\beta} \rVert ^2 _2 + \lambda \sum^p _{k=1} |\beta_k |\right)$, for a given penalty parameter $\lambda$. The LASSO inherently performs variable selection as the absolute value penalty induces sparsity as it shrinks some of the values in  $\bm{\beta}^\lambda$ all the way to zero.

The goal of stability selection is to estimate the set $\mathcal{S}$, dependent on the stable threshold parameter $\pi$, where variables are considered to be signal variables if their selection probabilities are greater than $\pi$ \citep{meinshausen_stability_2010}. When the variable selection algorithm requires a regularisation parameter, we recognise that the estimated set $\hat{\mathcal{S}}$ is a function of $\lambda$, and write $\hat{\mathcal{S}}^\lambda = \{ k: \hat{\beta}_k (\lambda) \neq 0,\ k= 1,\dots,p\}$ which is the set of selected variables for a given $\lambda$. Then, let $\hat{\mathcal{S}}^\Lambda = \bigcup_{\lambda \in \Lambda} \hat{\mathcal{S}}^\lambda$ be the set of selected variables for all values in a suitably chosen range of $\lambda \in \Lambda =  \{\lambda_1, \ldots, \lambda_t\}$. For each $\lambda \in \Lambda$, the variable selection algorithm produces an estimate $\hat{\mathcal{S}}^\lambda$ of the true set $\mathcal{S}$. The aim is to identify $\hat{\mathcal{S}}^\lambda$ such that there is a high probability of $\hat{\mathcal{S}}^\lambda$ being identical to the true but unknown set $\mathcal{S}$. Formally, we aim to achieve consistent variable selection, i.e.\ $\P(\hat{\mathcal{S}}^\lambda = \mathcal{S}) \rightarrow 1$ as $n \rightarrow \infty$. \citet{meinshausen_stability_2010} demonstrate that under general conditions, consistent variable selection is indeed achieved with stability selection. For the LASSO, the irrepresentable condition is required \citep{zhao_lasso_2006}. If this condition is not satisfied, the randomised LASSO can be used instead, which is consistent regardless \citep{meinshausen_stability_2010}.

Let $\mathcal{I}$ be a random subsample of $\{1,\dots,n\}$, of size $\lfloor \frac{n}{2}\rfloor$, drawn uniformly and without replacement. Let $\hat{\mathcal{S}}^\lambda(\mathcal{I})$ represent the selected variables given a particular subsample $\mathcal{I}$ and regularisation parameter $\lambda$.  The probability that the selected set $\hat{\mathcal{S}}^\lambda (\mathcal{I})$ contains a given set $\mathcal{K} \subseteq \{1,\dots,p\}$, is denoted as $\hat{\Pi}^\lambda _\mathcal{K}  = \P (\mathcal{K}\subseteq \hat{\mathcal{S}} ^\lambda (\mathcal{I}))$. Given the random subsample $\mathcal{I}$, let $q_\Lambda = \E(|\hat{\mathcal{S}}^\Lambda (\mathcal{I})|)$ be the average number of selected variables across all $\lambda \in \Lambda$. These probabilities are empirically estimated by observing the proportion of time a variable was selected over $B$ repeats of random subsamples. In the case of L1 regularised regression, a variable is considered selected when the estimated coefficient is non-zero.

For a stable threshold cut-off parameter  $\pi\in(0,1)$  and a set of regularisation parameters $\Lambda$, the set of stable variables is defined as $\hat{\mathcal{S}}^{\text{stable}} = \{k :\max_{\lambda \in \Lambda} (\hat{\Pi}^\lambda _k )\geq \pi,\ k=1,\dots,p\}$. The set of chosen stable variables $\hat{\mathcal{S}}^{\text{stable}}$ depends on the chosen threshold, $\pi$ whereby a variable is considered part of the stable set if the variable is selected at least $\pi \times 100$\% of the time from a variable selection algorithm. 

With the estimated stable variables in $\hat{\mathcal{S}}^{\text{stable}}$, we define $V$ as the number of falsely selected variables, $V = |\mathcal{N}\cap \hat{\mathcal{S}}^{\text{stable}}|$, where falsely selected variables are noise variables which are selected as stable. \citet{meinshausen_stability_2010} stated that in their empirical research, varying $\pi$ did not greatly vary the set of stable variables for $\pi$ between 0.6 and 0.9. However, we will show in Section \ref{sec:simulation} that choosing $\pi$ has an important role in determining the set of signal predictors, particularly in high-dimensional and low signal settings.  

Building on this foundation, our major contribution is automatic calibration of the stable threshold cut-off parameter $\pi$, inspired by the use of the elbow method in scree plots.

\section{Methodology}\label{chap:methodology}
The scree plot is commonly defined as a line plot displaying the eigenvalues for factors or principal components in principal component analysis. While this is the general use case for a scree plot, it does not necessarily need to follow this definition. We redefine a scree plot as a line plot that visualises the relationship between a parameter and a metric, such that an inflection point or ``elbow'' is able to be identified. Changing the parameter any further would then add little or no value. This estimated elbow is then considered the optimal value of the parameter of interest.

Although the elbow method appears straightforward, its implementation relies on a trained user to make informed decisions, as the lack of objectivity becomes particularly pronounced in more complex scree plots. Moreover, when the number of proposed parameters is large, static graphs may provide limited information; hence, the location of the elbow is difficult to determine.

In the application of scree plots for stability selection, we are implicitly optimising $\pi$, through the estimation of the number of variables in the stable set $\hat{\mathcal{S}}^{\text{stable}}$. The metric we are comparing the parameter $\pi$ against is the maximum variable selection probability $\max_{\lambda \in \Lambda} (\hat{\Pi}^\lambda _k ),\ k = 1,\dots,p$. With the help of the scree plot, only the variables with the highest selection probabilities should be selected, such that including additional variables could lead to overfitting or superfluous variable selections.

As discussed previously, visual approaches to finding the elbow rely on heuristic and loosely defined rules, sparking debate over whether the method should be used \citep{shi_quantitative_2021,yuan_research_2019}. To counteract this subjectivity, we modify the works of \citet{zhu_automatic_2006} who proposed a statistical approach to determine the elbow point without the need for visualisation. This procedure provides an objective estimate, which does not require tuning parameters, allowing complicated scree plots to be easily and automatically analysed.

Later in this section, we present EATS, a modified version of the algorithm proposed in \citet{zhu_automatic_2006}, optimised for stability selection. Our modified algorithm enhances the method by incorporating information through the estimation of noise variables' selection probabilities. To begin, we present ATS, the motivation and background that led to the development of EATS.

\subsection{Automatic Threshold Selection}\label{sec:ats}

Define a variable's selection frequency as the maximum number of occurrences for all $\lambda \in \Lambda$, where a variable was selected using a variable selection algorithm, over $B$ runs in the stability selection procedure. Let $\hat{d}_{\sigma(j)} = \max_{\lambda \in \Lambda} (\hat{\Pi}^\lambda _{\sigma(j)} ) $ be the $j$th ordered empirical selection probability, where $\sigma(j)$ gives the index of the $j$th ranked selection frequency in the original variable ordering. For ease of notation let $\hat{d}_j = \hat{d}_{\sigma(j)}$, but the mapping between the ordered empirical selection probability and original variable position is implicitly maintained. Define $\Delta$ as the non-increasing set of empirical selection probabilities for all variables $k \in \{1,\ldots, p\}$ over $B$ runs, $\Delta = \{\hat{d}_{1}, \hat{d}_2,\dots, \hat{d}_p:\ 1 \geq \hat{d}_1 \geq \hat{d}_2 \geq \cdots \geq \hat{d}_p \geq 0\}$. In other words, $\Delta$ is the non-increasing ordered set of variable selection probabilities, as obtained from the stability selection procedure.

For $w = 1,2,\dots, p-1$, if there exists an elbow at index $w$, we can think of $\bm{\delta}_1 = \{\hat{d}_1,\hat{d}_2,\dots,\hat{d}_w\}$ and $\bm{\delta}_2 = \{\hat{d}_{w+1},\hat{d}_{w+2},\dots,\hat{d}_p\}$ as samples from two distinct distributions, with the same functional form: $f(\bm{\delta}_1;\bm{\theta}_1), f(\bm{\delta}_2;\bm{\theta}_2)$. Here, $\bm{\theta}_1$ and $\bm{\theta}_2$ contain all relevant parameters for their respective distributions and are considered nuisance parameters. Following \citet{zhu_automatic_2006}, we assume $f(\bm{\delta}_i;\bm{\theta}_i), i= 1,2$ follows a normal distribution which enables the computation of the profile log-likelihood for each $w = 1, 2, \ldots, p-1$ given as 
\begin{align}\label{eq:MLE}
    l(w) = (-p-w)\log(\hat{\sigma} \sqrt{2\pi}) - \frac{1}{2\hat{\sigma}^2} \left(\sum^w _{i=1} (\hat{d}_i - \hat{\mu}_1 )^2 + \sum^p _{m=w+1} (\hat{d}_m - \hat{\mu}_2)^2 \right),
\end{align} 
where $\hat{\mu}_1$ and $\hat{\mu}_2$ are the maximum likelihood estimates (MLEs) for the sample mean of $\bm{\delta}_1$ and $\bm{\delta}_2$ respectively, and $\hat{\sigma}^2$ is the MLE for the common pooled scale parameter. Furthermore, we define $\Omega$ as the set of profile log-likelihoods for each candidate elbow, $\Omega = \{l(1),\dots,l(p-1)\},\ w = 1,\dots,p-1$. The candidate elbow $w$ that maximises $\Omega$ is then the estimated elbow point, $\hat{w} = \argmax_{w} \Omega$. The estimated elbow's selection probability is considered the estimated stable threshold parameter and can be represented as the $\hat{w}$th index in $\Delta$, $\hat{\pi}(\hat{w}) = \Delta [\hat{w}]$. Then we can state that all variables with a maximum selection probability greater than or equal to $\hat{\pi} (\hat{w})$ are considered part of the stable set of variables, $\hat{\mathcal{S}}^{\text{stable}}$. That is, with $\pi$ as $\hat{\pi}(\hat{w})$, $\hat{\mathcal{S}}^{\text{stable}} = \{k :\max_{\lambda \in \Lambda} (\hat{\Pi}^\lambda _k )\geq \pi; k=1,\ldots,p\}$.

\subsection{Exclusion Probability Threshold}
Our empirical studies in Section \ref{chap:numerical} indicate that applying ATS to a large number of variables and/or low signal-to-noise ratios will yield a larger elbow index, and therefore risks overselecting variables. A potential reason for this characteristic is that the noise variables with small selection probabilities possess excessive weight in the ATS procedure. As a solution, we recommend an initial filtering process to reduce the number of candidate elbows by generating a set of noise variables that are uncorrelated with the outcome variable. 

Randomly reorder the rows in $\bm{Z} =(\bm{X} ,\bm{y})$ and push each entry in $y_i$ to $y_{i+1}$ for $i =1,\dots,n-1$, while setting $y_{1} = y_n$. Then $\bm{X}^*, \bm{y}^* $ is the shuffled counterpart of $\bm{X}$, $\bm{y}$ respectively, such that no observations are matched with their true outcome. Denote $\Delta^*$ as the non-increasing set of variable selection probabilities for the dataset $\bm{X}^*$ and $\bm{y}^*$ where $\hat{d}^* = \max_{\lambda \in \Lambda} (\hat{\Pi}^{\lambda*})$ and $\Delta^* = \{\hat{d}_{1}^*, \hat{d}_2^*,\dots, \hat{d}_{p}^*:\ 1 \geq \hat{d}^*_1 \geq \hat{d}^*_2 \geq \cdots \geq \hat{d}_{p}^*\}$, analogous to the parameters $\hat{d},\hat{\Pi}^\lambda,$ and $\Delta$. The exclusion probability threshold $\eta$ is then defined as the $95$th percentile of $\Delta^*$, i.e.\ $\eta = \hat{F}^{-1} _{\Delta^*} (0.95).$

Define $\tilde{p}$ as the total number of variables with selection probabilities greater than the exclusion probability. Mathematically, $\tilde{p}$ = $\max \{j\ |\ \hat{d}_j \geq \eta,\ j = 1,\dots, p\}$, the largest $j$ such that $\hat{d}_j \geq \eta$. Also define $\tilde{\Delta} \subseteq \Delta$ as the non-increasing empirical variable selection probabilities for all $\hat{d}_j \geq \eta$, $j = 1, \dots, \tilde{p}$. Then,
\begin{align*}
\tilde{\Delta} = \{\hat{d}_{1}, \hat{d}_2,\dots, \hat{d}_{\tilde{p}}:\ 1 \geq \hat{d}_1 \geq \hat{d}_2 \geq \cdots \geq \hat{d}_{\tilde{p}} \geq \eta\}.
\end{align*}

The exclusion probability threshold allows $(\bm{X}^*,\bm{y}^*)$ to retain the dependency structure of $\bm{X}$ whilst discarding any potential signal. This produces $p$ known noise variables. These selection probabilities of the known noise variables allow us to estimate the potential selection probabilities for the unknown noise variables in $(\bm{X},\bm{y})$. The assumption here is that we expect variables with a smaller selection probability than $\eta$ to be noise, and should not be considered in our ATS procedure. After eliminating the estimate noise variables, we are left with the filtered set of candidate elbows $\tilde{\Delta}$.

We see in our simulated results in Section \ref{chap:numerical} that incorporating such information reduces the chance of selecting spurious variables and improves overall performances in certain criteria, especially when $p>n $. A detailed summary of EATS can be seen in Algorithm \ref{alg:ats}.

\let\oldnl\nl
\newcommand{\nonl}{\renewcommand{\nl}{\let\nl\oldnl}}
\SetNlSty{texttt}{(}{)}
\SetAlgoVlined
\DontPrintSemicolon
\begin{algorithm}
\KwIn{$\bm{X},\bm{y}, \Lambda, B$} 
\tcp{\textbf{\textrm{Stability Selection Step}}}
$\hat{\Pi}^{\lambda} \gets \text{ Do stability selection on } (\bm{X},\bm{y},\Lambda, B)$\;
$\hat{d} _j\gets \max_{\lambda \in \Lambda} (\hat{\Pi}^{\lambda}_j) \text{ for }j\in\{1,\dots,p\}$\;
$\Delta \gets \{\hat{d}_{1}, \hat{d}_2,\dots, \hat{d}_{p}:\ 1 \geq \hat{d}_1 \geq \hat{d}_2 \geq \cdots \geq \hat{d}_{p}\}$\;

\vspace{1em}
\tcp{\textbf{\textrm{Exclusion Probability Threshold Step}}}
$\bm{Z}^* = (\bm{X}^*,\bm{y}^*) \gets \text{Randomly reorder the rows in }\bm{Z} = (\bm{X},\bm{y})$\;
$y^* _1 \gets y_n$\;
\For{$i=1$ \KwTo $n-1$}{
$y^* _{i+1} \gets y_{i}$\;
}

$\hat{\Pi}^{\lambda*} \gets $ Do stability selection on $(\bm{X}^*, \bm{y}^*, \Lambda, B)$\;
$\hat{d}^* _j\gets \max_{\lambda \in \Lambda} (\hat{\Pi}^{\lambda *}_j) \text{ for }j\in\{1,\dots,p\}$\;
$\Delta^* \gets \{\hat{d}_{1}^*, \hat{d}_2^*,\dots, \hat{d}_{p}^*:\ 1 \geq \hat{d}^*_1 \geq \hat{d}^*_2 \geq \cdots \geq \hat{d}_{p}^*\}$\;
$\eta \gets \hat{F}^{-1} _{\Delta^*} (0.95)$\;
\vspace{1em}
\tcp{\textbf{\textrm{Automatic Threshold Selection Step}}}

$\tilde{p} \gets \max \{j\ |\ \hat{d}_j \geq \eta,\ j = 1,\dots, p\}$\; 
$\tilde{\Delta} \gets \{\hat{d}_{1}, \hat{d}_2,\dots, \hat{d}_{\tilde{p}}:\ 1 \geq \hat{d}_1 \geq \hat{d}_2 \geq \cdots \geq \hat{d}_{\tilde{p}} \geq \eta\}$\;
$\Omega \gets \emptyset$\;
\For{$w = 1$ \KwTo $\tilde{p} - 1$}{
    $\hat{\mu}_1 \gets \frac{1}{w} \sum_{i=1} ^w  \tilde{\Delta}[i]$\;
    $\hat{\mu}_2 \gets  \frac{1}{{\tilde{p}}-w}\sum_{j = w + 1} ^{\tilde{p}}  \tilde{\Delta}[j]$\;

    $\hat{\sigma}^2 \gets \frac{1}{\tilde{p}-2}(\sum_{i= 1} ^w ( \tilde{\Delta}[i] - \hat{\mu}_1 )^2 + \sum_{j = w + 1} ^ {\tilde{p}} ( \tilde{\Delta}[j] - \hat{\mu}_2)^2 )$\;

    $\Omega[w] \gets \sum_{i=1} ^w \log(f( \tilde{\Delta}[i],\hat{\mu}_1, \hat{\sigma}^2))\ + \sum^{\tilde{p}} _{j = w+ 1} \log (f( \tilde{\Delta}[j],\hat{\mu}_2,\hat{\sigma}^2))$\;
}
$\hat{w} \gets \argmax_{w=1,2,\dots,\tilde{p}-1} \Omega$\;
$\hat{\pi}(\hat{w}) \gets \Delta [{\hat{w}}]$\;
\vspace{1em}
\KwRet{$\hat{\pi}(\hat{w})$}
\caption{EATS}\label{alg:ats}
\end{algorithm}

\subsection{Error Control}\label{sec:errcontrol}

\citet{meinshausen_stability_2010} provide the mathematical framework for controlling false discovery error rates in the context of stability selection. In this section, we demonstrate how to implement EATS while maintaining error control. 

The error control bound can be represented through three tunable parameters $\E(V), q_\Lambda$, and $\pi$. First define $V = |\mathcal{N} \cap \hat{\mathcal{S}}^{\text{stable}} |$ as the number of falsely selected variables with stability selection for a given stable threshold parameter, $\pi$. Then $\E(V)$ is the expected number of falsely selected variables. Let $q_\Lambda = \E(|\hat{\mathcal{S}}^\Lambda (\mathcal{I})|)$ be the average number of selected variables for all $\lambda \in \Lambda$ and a given subsample $\mathcal{I}$.

Following \citet{meinshausen_stability_2010}, we are required to assume that the distribution of $\{\mathbbm{1}_{k\in \hat{\mathcal{S}}^\lambda},k\in \mathcal{N}\}$ is exchangeable for all $\lambda \in \Lambda$ with $\mathbbm{1}$ denoting the indicator function. We are also required to assume that the original procedure is not worse than random guessing, i.e. for any $\lambda \in \Lambda$,
            $\frac{\E(|\mathcal{S}\cap \hat{\mathcal{S}}^\Lambda|)}{\E (|\mathcal{N} \cap \hat{\mathcal{S}}^\Lambda |)} \geq \frac{|\mathcal{S}|}{|\mathcal{N}|}.$
    When $\pi \in (\frac{1}{2},1)$, $\E(V)$ is then bounded by,
\begin{align}\label{eq:err}
        \E (V) \leq \frac{1}{2\pi - 1} \frac{q^2 _\Lambda }{p}.
\end{align}
Fixing the error bound stated in \eqref{eq:err} requires defining two of the three parameters $\E(V), q^2 _{\Lambda}$ and $\pi$, as the third can be computed given the first two. Since EATS estimates $\pi$, we only require specifying either the desired error tolerance $\E(V)$, or the average number of selected variables over $\Lambda$, $q_{\Lambda}$. While either can be defined, we find that it is more intuitive to specify the error tolerance. Regardless, the third parameter can be solved via the inequality $q^2_\Lambda \geq \E(V)(2\hat{\pi}(\hat{w}) - 1)p$ for $q^2 _\Lambda$ or using \eqref{eq:err} for $\E(V)$.

For more difficult variable selection problems such as a high-dimensional setting, we may see that even signal variables exhibit selection probabilities less than 0.5, and hence $\hat{\pi}(\hat{w}) < 0.5$. For this, a user may restrict $\pi = \max\{\hat{\pi}(\hat{w}), 0.5\}$ or utilise Complementary Pairs Stability Selection (CPSS) which provides an error bound for threshold values of less than 0.5 \citep[Theorem 1]{shah_variable_2013}.

\section{Numerical Results}\label{chap:numerical}
In this section, we applied EATS to both artificial and real datasets along with a simulated error control study. We considered multiple stable threshold parameters in our simulation study; ATS-selected $\pi$, EATS-selected $\pi$, and one static value of $\pi$ which fall at the midpoint of the suggested range given by \citet{meinshausen_stability_2010}: $0.75$. The variable selection algorithm used in the stability selection procedure was the LASSO for all stable threshold parameter estimation methods. Throughout our simulation study, we also benchmarked stability selection against the LASSO (without stability selection), knockoff, and the smoothly clipped absolute deviation penalty (SCAD) \citep{barber_controlling_2015, fan_variable_2001, tibshirani_regression_1996}. The regularisation term $\lambda$ for LASSO and SCAD was the value that minimised the 10-fold cross-validation error. The knockoff target false discovery parameter was held at 0.1, the default parameter setting as in \citet{patterson_knockoff_2022}. Additionally, datasets with $p>500$ utilised the approximate semidefinite programming knockoff (ASDP), instead of the regular SDP, as specified by default in \citet{patterson_knockoff_2022}.

For all stability selection methods (ATS, EATS, and static), the original subsampling method of \citet{meinshausen_stability_2010} was used, and the error bound $\E(V)$ was held at $5$. We reported the Matthews correlation coefficient (MCC) and the number of selected variables as the main forms of evaluation across 1000 simulation runs for each setting \citep{matthews_comparison_1975}. Appendix \ref{app:sims} includes a variety of supplementary material, such as the empirical distribution of the exclusion probability threshold $\eta$ and EATS-estimated $\hat{\pi}(\hat{w})$.

All computation was performed using R \citep{r_core_team_r_2024}. Stability selection was conducted through the \textit{stabsel} package \citep{hofner_stabs_2021}. The knockoff was computed through the \textit{knockoff} package, SCAD through the \textit{ncvreg} package and LASSO through the \textit{glmnet} package \citep{breheny_coordinate_2011, friedman_regularization_2010,patterson_knockoff_2022}. The ATS and EATS implementation code along with some example code for this simulation study can be found at {https://github.com/MartinHuangR/Automatic-Threshold-Selection}

\subsection{Artificial Data} \label{sec:simulation}
All artificial data simulation studies used similar data generation processes, differing only in the dimensionality of the design matrix $\bm{X}$ and response variable $\bm{y}$. We generated the $i$th predictor vector from $\bm{x}_i \sim N(\bm{0}, \bm{\Sigma})$, where $(\bm{\Sigma})_{ij} = 0.5 ^{|i - j|}$ was the covariance matrix with a Toeplitz design. The coefficients of signal variables $\bm{\beta}_\mathcal{S}$ were randomly sampled from $\{-3, -2, -1, 1, 2, 3\}$ and $|\bm{\beta}_\mathcal{S}|$ denotes the cardinality of the set $\mathcal{S}$, i.e.\ the number of signal variables. Furthermore, by definition, the coefficients of noise variables $\bm{\beta}_{\mathcal{N}}$ were zero. We generated the response through $\bm{y} = \bm{X} \bm{\beta} + \bm{\epsilon}$, where $\bm{\beta} = [\bm{\beta}_\mathcal{S}^\top,\bm{\beta}_\mathcal{N}^\top]^\top$, and $\bm{\epsilon} \sim N(\bm{0},\sigma^2\bm{I})$. We also considered four signal-to-noise ratios (SNR): 0.5, 1, 2, and 3, such that $\text{SNR} = \lVert \bm{X} \bm{\beta}\rVert ^2 _2 /n\sigma^2$. Since $\bm{X}$ is fixed and $\bm{\beta}$ is artificially generated, $\sigma$ is chosen to achieve the desired SNR. 

Our artificial simulation study also examined four combinations of $n$, $p$, and $|\bm{\beta}_\mathcal{S}|$:
\begin{alignat*}{7}
&\text{(I)}  &&\quad n = 20,  && \quad p &&= 1000,  && \quad |\bm{\beta}_\mathcal{S}| &&= 2\\
&\text{(II)} &&\quad n = 100,  && \quad p &&= 500,   && \quad |\bm{\beta}_\mathcal{S}| &&= 10\\
&\text{(III)} &&\quad n = 200,  && \quad p &&= 200,   && \quad |\bm{\beta}_\mathcal{S}| &&= 20\\
&\text{(IV)} &&\quad n = 500,  && \quad p &&= 100,   && \quad |\bm{\beta}_\mathcal{S}| &&= 20 
\end{alignat*}

For all simulation settings, the randomly generated coefficients that were sampled from $\{-3, -2, -1, 1, 2, 3\}$ can be found in Table \ref{tab:coefficients}.

Using the same structure, we later investigated how the characteristics of ATS and EATS change when $n$ is fixed and $p$ varies, and vice versa.

\begin{figure}[htbp]
    \centering
    \includegraphics[width = 0.9\textwidth]{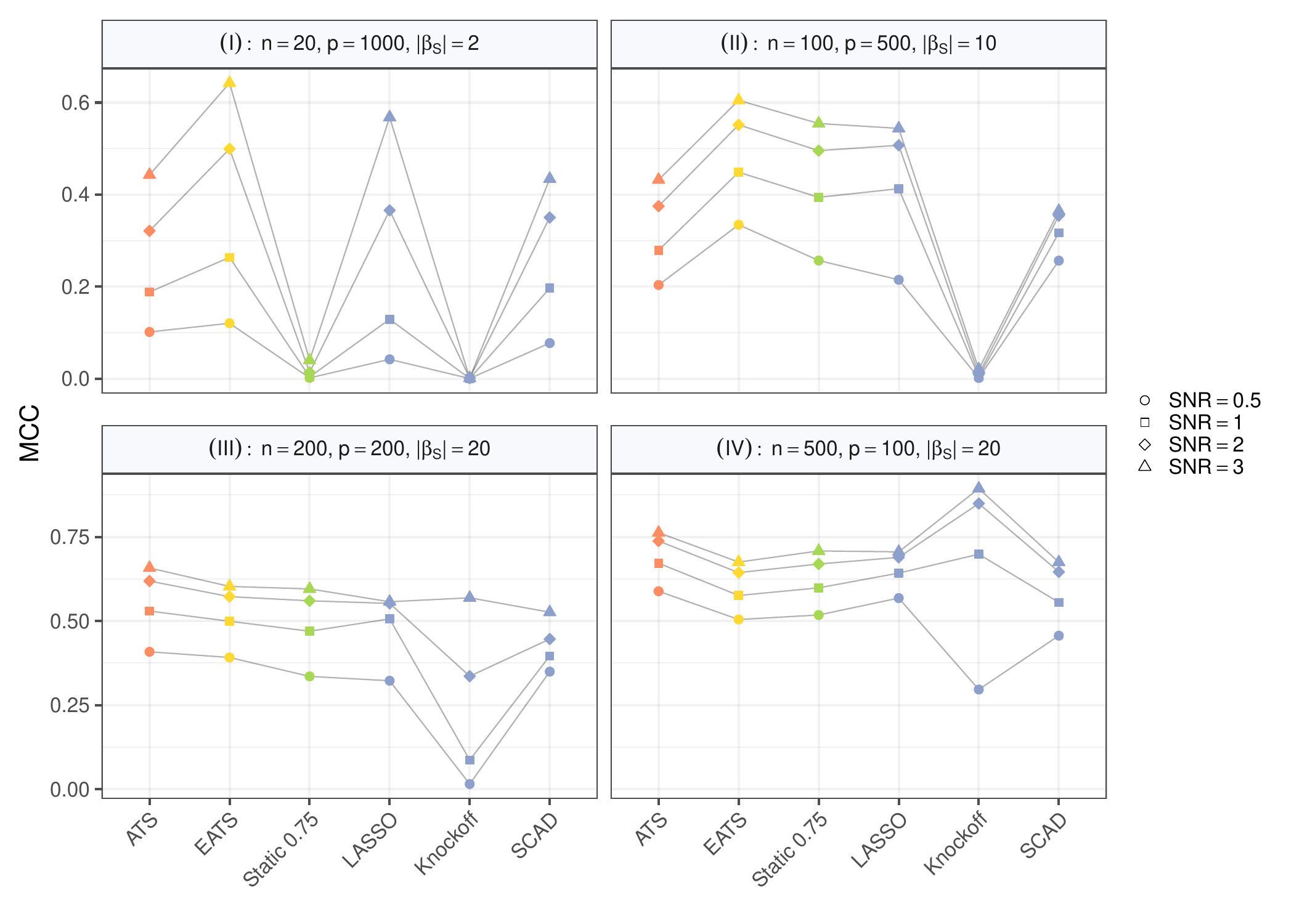}
    \caption{MCC score for simulation study settings (I) - (IV) with varying SNR across the different methods.}
    \label{fig:mcchard}
\end{figure}

The simulation results shown in Figure \ref{fig:mcchard} reveal that the ATS-estimated values of $\pi$ produces MCCs that are either comparable to or exceed all competing methods, for settings (III) and (IV). On the other hand, EATS outperforms all methods in setting (I) and (II). In setting (I) where the data is high-dimensional, the static value fails to regularly detect the low selection probabilities of signal variables, particularly as the SNR values is small.

\begin{figure}[htbp]
    \centering
    \includegraphics[width = 0.9\textwidth]{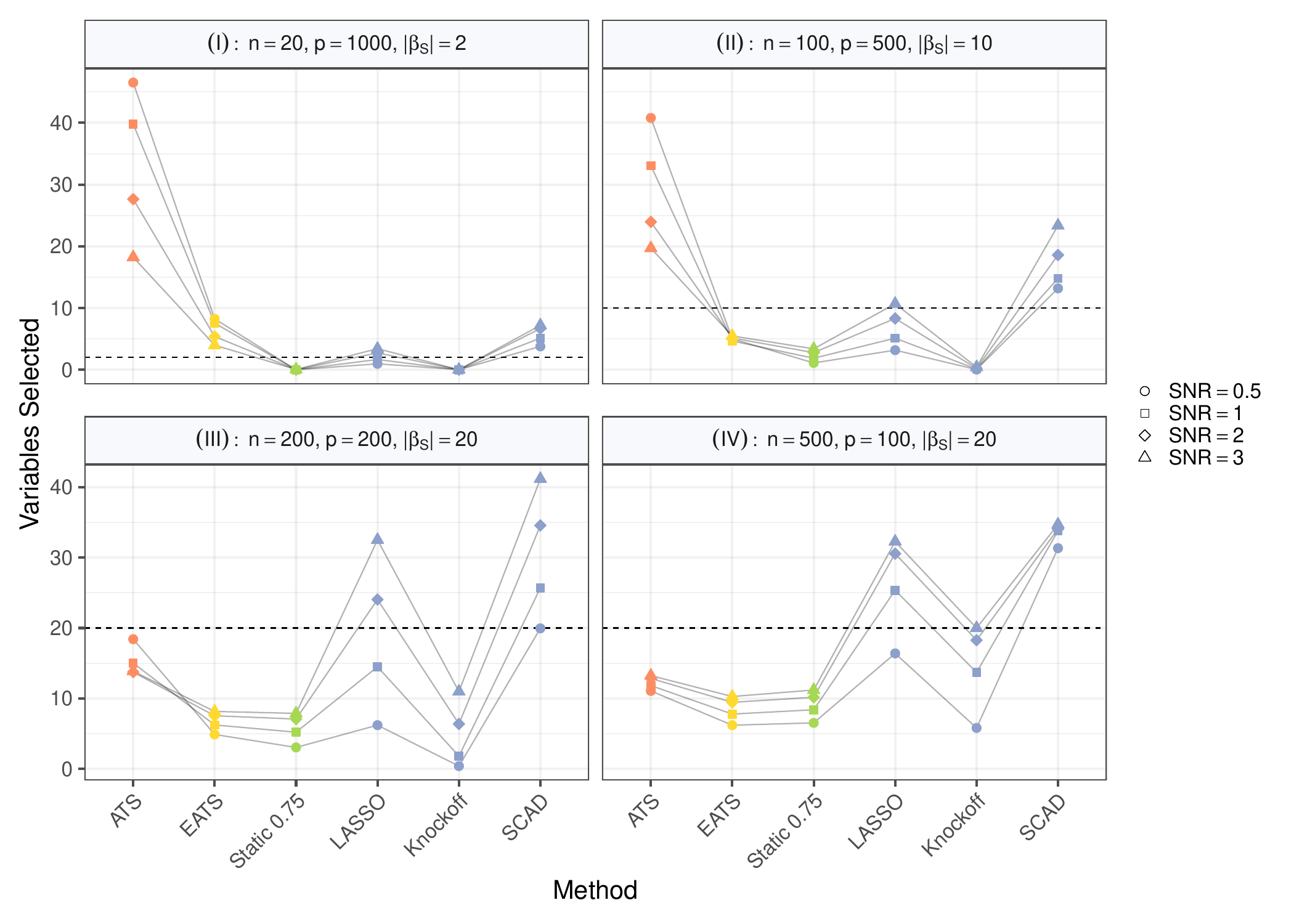} 
    \caption{Number of selected variables for simulation study settings (I) - (IV) with varying SNR. The dashed line denotes the number of active variables $|\bm{\beta}_S|$.}
    \label{fig:nplothard}
\end{figure}

While ATS and EATS do not substantially increase the performance of stability selection relative to the static value, they reduce the chance of selecting an erroneous $\pi$. In all four simulation settings, our analysis (not shown) indicates that selecting a value of $\pi$ greater than $0.75$ would subsequently underselect, and hence reduce the MCC. Furthermore, Figure \ref{fig:mcchard} demonstrates that EATS considerably outperforms ATS in scenarios where $p > n$, such as in settings (I) and (II), making it the preferred method for high-dimensional data. The improved MCCs achieved by EATS can be attributed to its use of the exclusion probability threshold $\eta$, which reduces the number of selected variables compared to ATS, as illustrated in Figure \ref{fig:nplothard}. 

In settings (III) and (IV) where $p \leq n$, ATS performs slightly better as the $\eta$ in EATS reduces too many candidate elbows, resulting in underselecting variables (Figures \ref{fig:etasims}, \ref{fig:pisims}). Even when EATS selects fewer variables than ATS, it still performs well relative to the static $\pi$.

\begin{figure}[htbp]
    \centering
    \includegraphics[width = 0.9\textwidth]{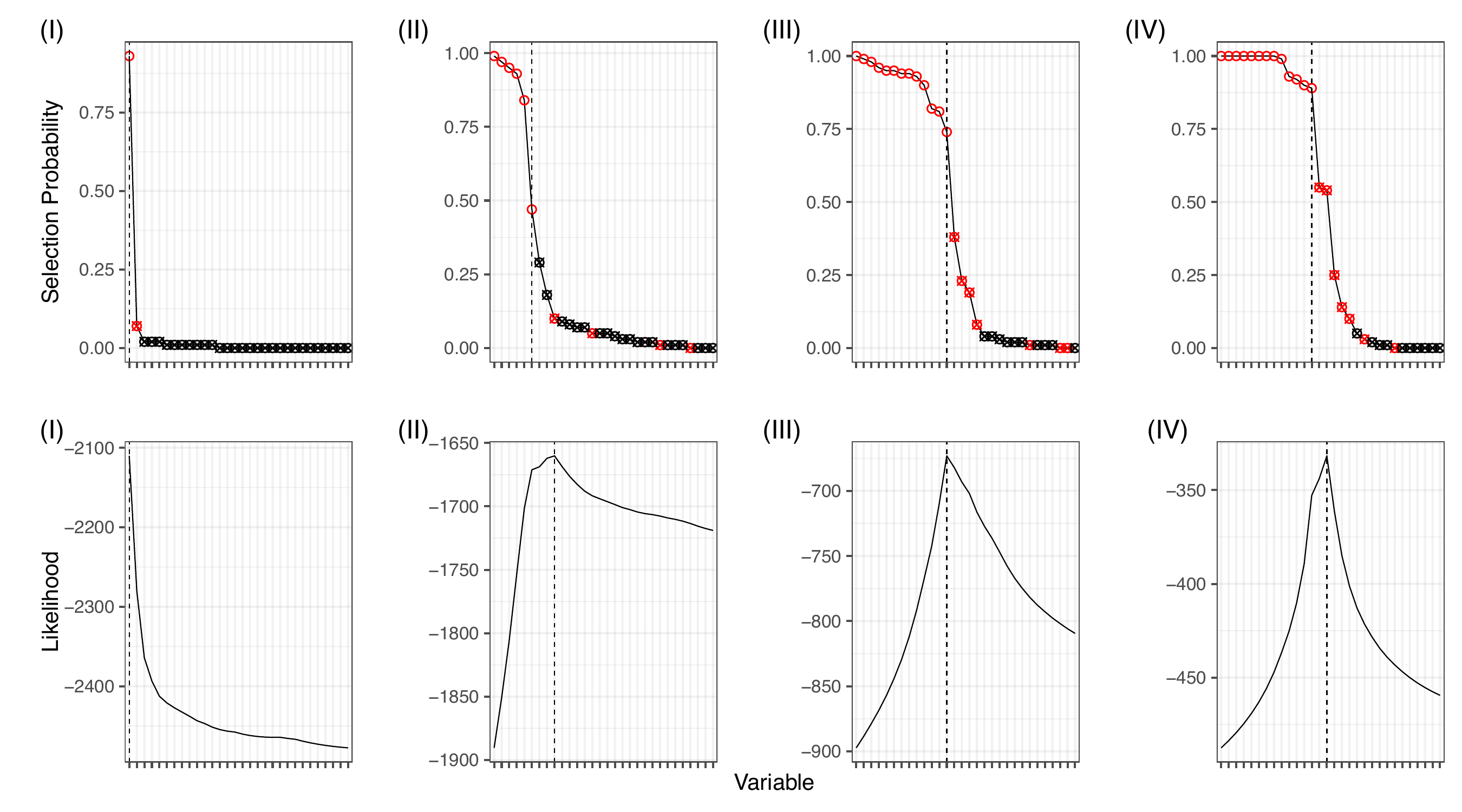}
    \caption{The top row shows scree plots for the selection probabilities of the $30$ most frequently selected variables. The circle shape indicates an EATS-selected variable and falls on the left of the elbow (dotted line). The circle-cross shapes indicate non-selected variables. The red (black) highlighted points indicate the true signal (noise) variables. In the likelihood plots (bottom row), the dotted lines display the index of the variable that maximises the likelihood function. The dataset settings (I) - (IV) are used from the artificial datasets with SNR $=3$. }
    \label{fig:elbowplot}
\end{figure}

For each setting, we investigated a single simulation regarding how the elbow is estimated through the corresponding likelihood function in Figure \ref{fig:elbowplot}. In all settings apart from setting (I), the elbow is not necessarily clear. For example, in setting (IV) we could consider a second elbow, around selection probability 0.5, which would yield two more correctly identified signal variables. However, including this second elbow would risk overselecting variables. Overall, the elbows selected by EATS align with our subjective assessment and function as intended.

\begin{figure}[htbp]
    \centering
    \includegraphics[width = 0.9\textwidth]{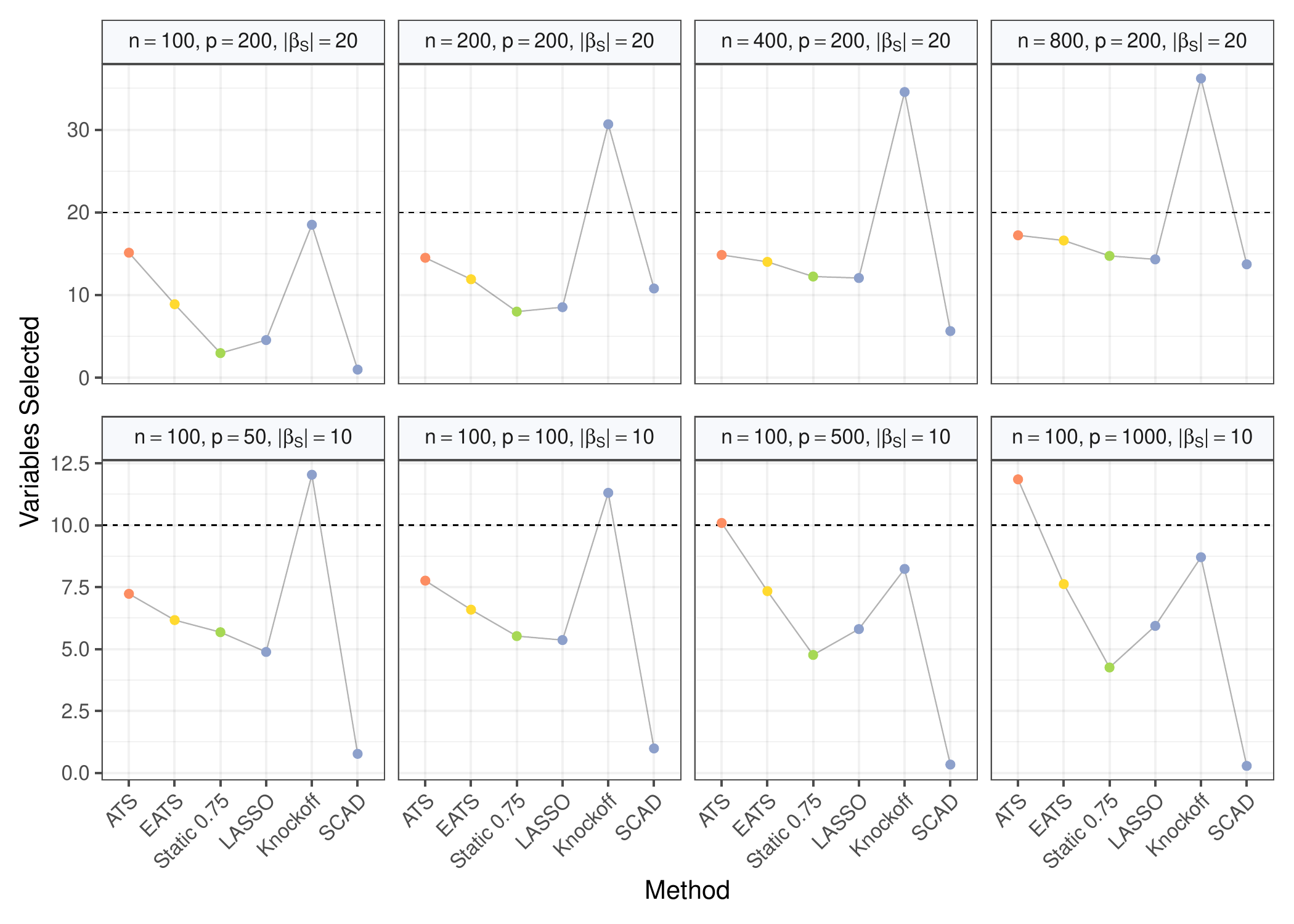}
    \caption{Influences in variables selected when varying $p$ and $n$ for a fixed $\text{SNR} = 3$.}
    \label{fig:II-IIIn}
\end{figure}

We considered the effect of increasing the number of predictor variables ($ p = 50, 100, 500, 1000$) while keeping the sample size fixed at $n = 100$. The specific coefficients can be found in Table \ref{tab:coefficientsNP}. Aligned with the results in Figure \ref{fig:nplothard}, Figure \ref{fig:II-IIIn} demonstrates that as $p$ increases, the number of selected variables also increases for both ATS and EATS. Although both methods select more variables as $p$ grows, EATS demonstrates greater robustness due to the exclusion probability threshold, which subsequently reduces the number of candidate elbows.

\begin{figure}[htbp]
    \centering
    \includegraphics[width = 0.9\textwidth]{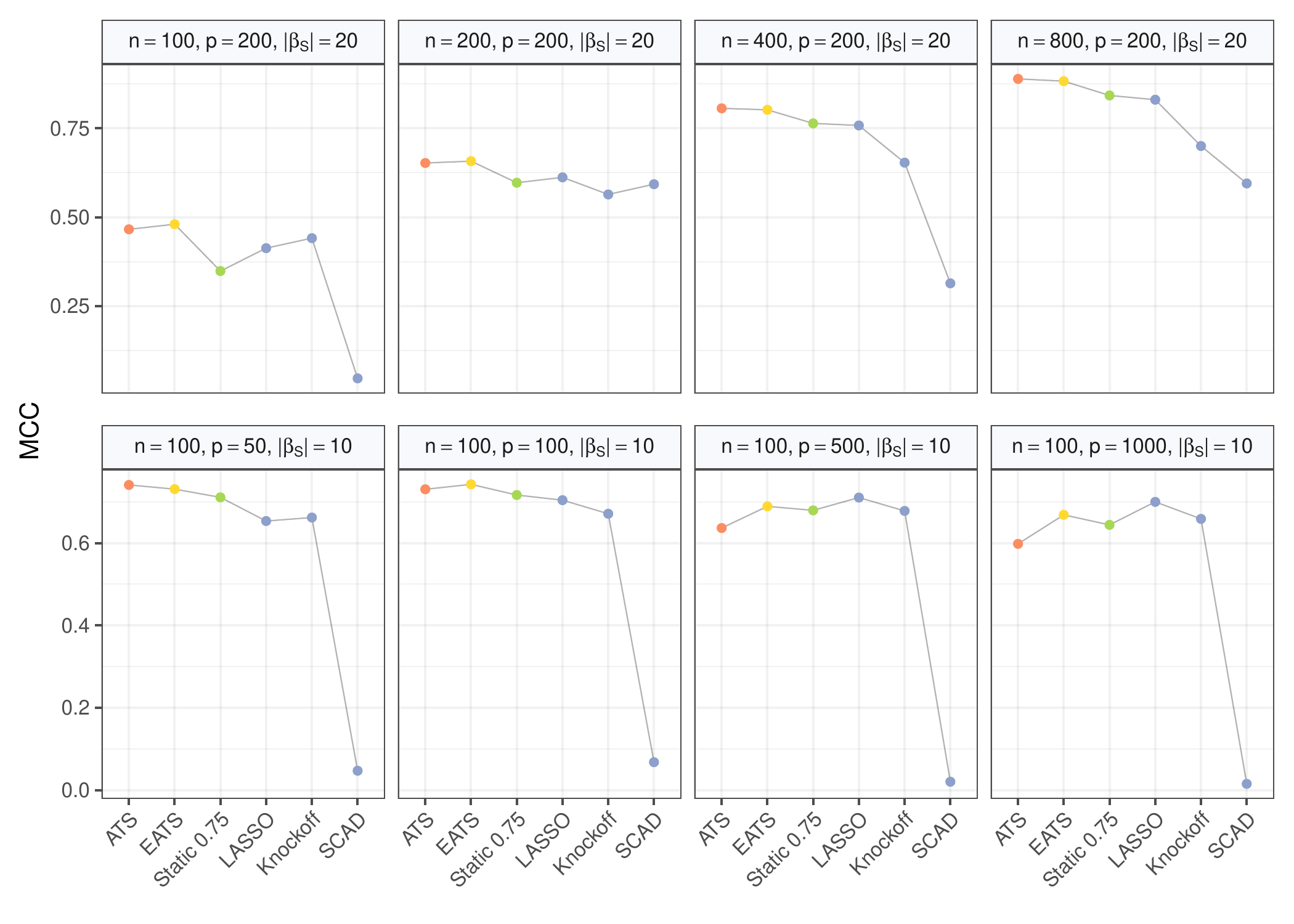}
    \caption{Influences in MCC when varying $p$ and $n$ for a fixed $\text{SNR} = 3$.}
    \label{fig:II-III}
\end{figure}

From Figure \ref{fig:II-III} and \ref{fig:II-IIIn}, when $p = 200$ is fixed, the MCC improves and the number of selected variables trends towards the number of active variables, as the sample size grows ($n = 100,200,400,800$). The only distinction in MCC performance between ATS and EATS occurs when $p$ is greater than $n$. For all other cases, the performances are comparable.

\begin{figure}[htbp]
    \centering
    \includegraphics[width = 0.9\textwidth]{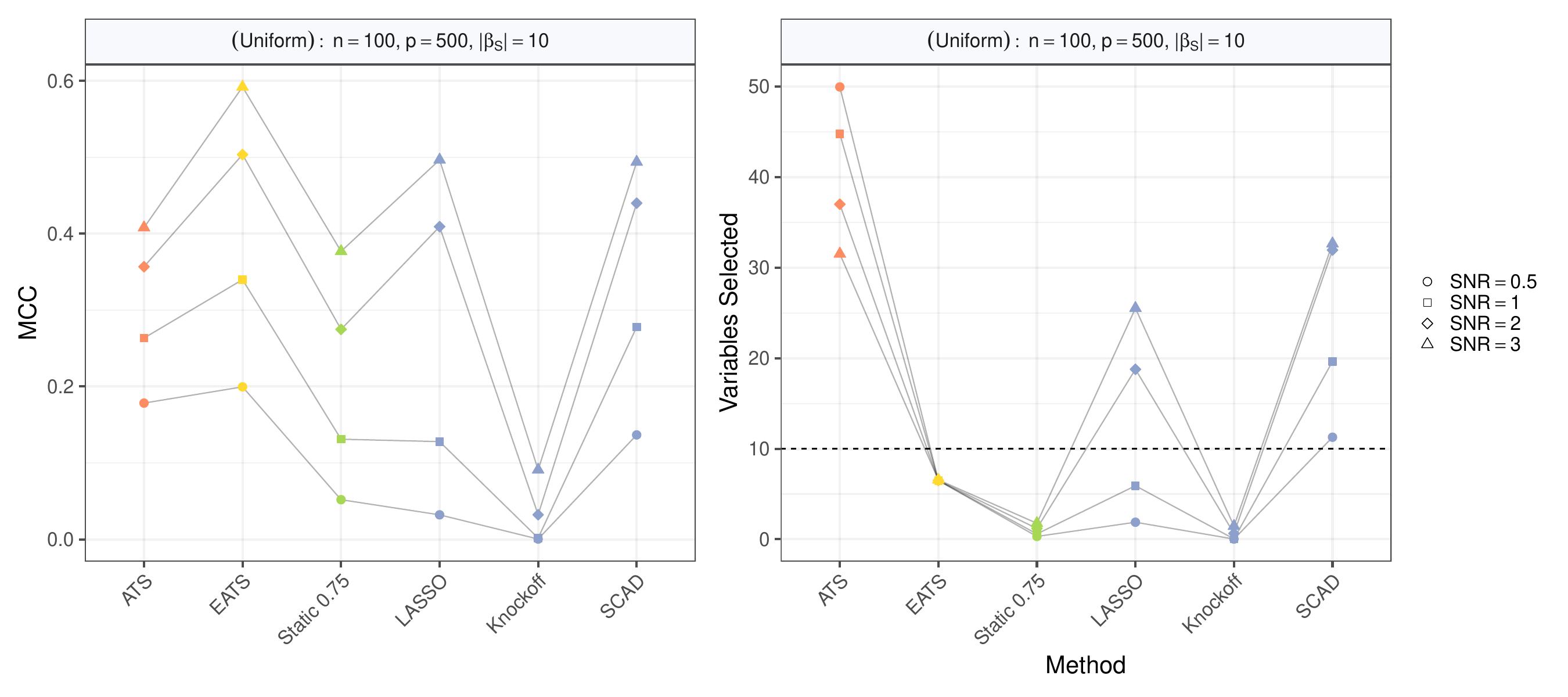}
    \caption{MCC and number of selected variables for simulation with data generated from a standard uniform with no correlation and varying SNR.}
    \label{fig:unif}
\end{figure}

\begin{figure}[htbp]
    \centering
    \includegraphics[width = 0.9\textwidth]{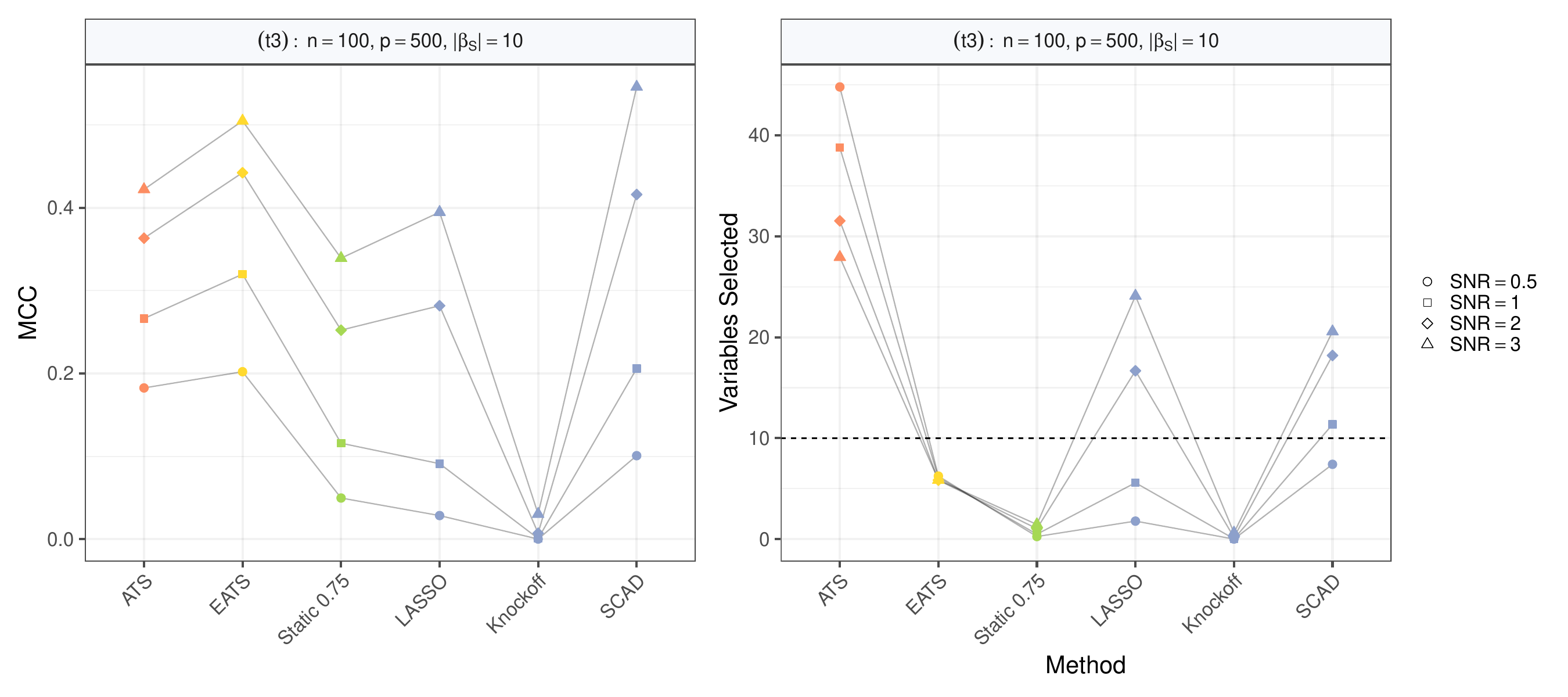}
    \caption{MCC and number of selected variables for simulation with data generated from a $t_3$ distribution with a Toeplitz matrix with the highest correlation of $0.2$ and varying SNR.}
    \label{fig:t3}
\end{figure}

We also considered settings with non-Gaussian design matrices. Figure \ref{fig:unif} shows results for standard uniform and uncorrelated predictors. Figure \ref{fig:t3} shows results for $t_3$-distributed predictors. Furthermore, these predictors were set to be lowly correlated with each other, following a Toeplitz covariance matrix with the highest correlation of $0.2$. In these two cases, EATS clearly outperforms the static value in MCC, while selecting the most appropriate amount of variables. In comparison to the other competitors, EATS is shown to be superior, apart from one case where the $t$-distribution is combined with SNR value of $3$.

\subsection{Plasma Proteomics}\label{sec:pro}

In this section we considered a plasma proteomics dataset as presented in \citet{hedou_discovery_2024} and \citet{rumer_integrated_2022}. We applied EATS to this dataset to demonstrate the application to a realistic high-dimensional dataset. The proteomics data was extracted from a clinical study of patients, undergoing non-urgent major abdominal colorectal surgery. The aim of the study was to analyse pre-operative blood samples to develop a model to predict patients at risk of post-operative surgical site infection. 

The dataset contained $n = 93$ patients, and $p = 722$ inflammatory proteins, and is accessible at {https://github.com/gregbellan/Stabl/}. We defined the plasma concentration to be our design matrix and generated the response with pre-defined signal variables. With the design matrix standardised at mean $0$ and standard deviation $1$, we set $\bm{y} = \bm{X}\bm{\beta} + \bm{\epsilon}$. We considered two simulation settings, where only the number of signal variables differ. The number of active variables was set to correspond to 5\% or 10\% of the sample size. The first setting had $|\bm{\beta}_\mathcal{S}| = 9$, $\bm{\beta} =  \{1,1,1,1,1,1,1,1,1,0,\dots,0\}$ and the second setting had $|\bm{\beta}_\mathcal{S}| = 4$, $\bm{\beta} =  \{1,1,1,1,0,\dots,0\}$. We also considered two SNR values, 1 and 3 with  $\bm{\epsilon} \sim N(\bm{0},\sigma^2\bm{I})$.

The application of the ASDP knockoff to the proteomics dataset proved computationally intensive, especially for repeated simulations. To reduce the computation cost of the ASDP knockoff, this section utilised the equi-correlated knockoff, as recommended by \citet{patterson_knockoff_2022}.  The equi-correlated knockoff was implemented via the \textit{knockoff} R package and is a computationally cheaper alternative than the regular knockoff, however possesses less statistical power \citep{candes_panning_2018, patterson_knockoff_2022}.

\begin{figure}[htbp]
    \centering
    \includegraphics[width = 0.9\textwidth]{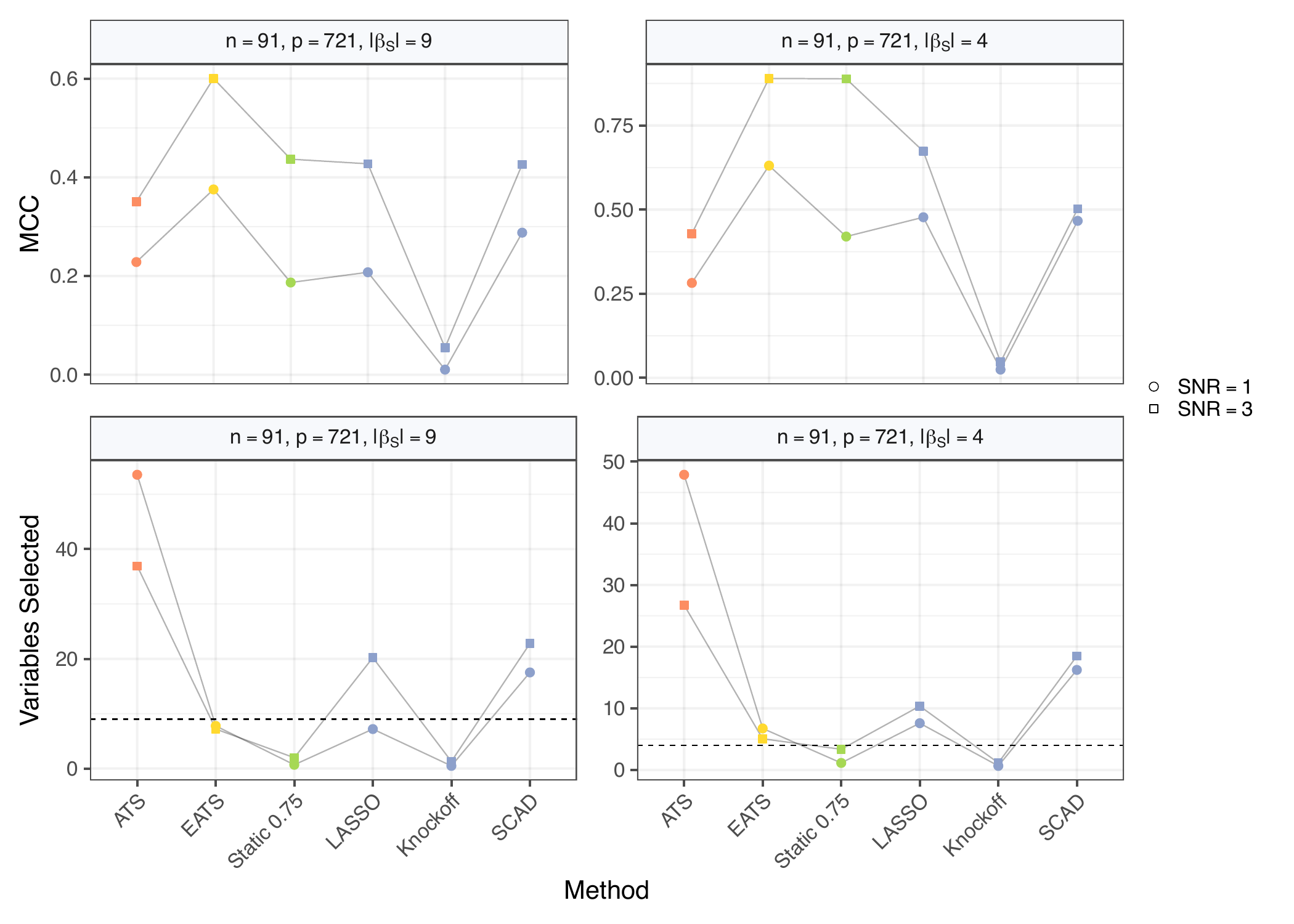} 
    \caption{Plasma proteomics ($n = 91, p = 721$) simulations for $|\bm{\beta}_{\mathcal{S}}| = 4$ and $9$ with SNR $= 1$ and $3$. The dotted line denotes the expected number of selected variables.}
    \label{fig:pro}
\end{figure}

As shown in Figure \ref{fig:pro}, an exclusion threshold substantially increases the MCC between ATS and EATS, no matter the number of active variables $|\bm{\beta}_\mathcal{S}|$ or SNR values. Furthermore, Figure \ref{fig:pro} demonstrates that ATS considerably overselects the total number of variables and is only reduced to an acceptable range when the exclusion probability threshold is introduced in EATS. Furthermore, EATS outperforms all other methods.

\subsection{Diabetes Study}

To evaluate the effectiveness of our methods when the number of predictors is smaller than the number of observations, we analysed a diabetes dataset from \citet{efron_least_2004}, obtained via the \textit{lars} R package \citep{efron_lars_2022}. The study originally included $10$ baseline variables, capturing various details on $n = 442$ diabetes patients. With the first $10$ variables being the baseline variables, the dataset also included $54$ other quadratic and interaction terms, totalling $p = 64$ variables.

The $p = 64$ candidate predictors acted as the design matrix and were used to generate a simulated response, based on a set of known signal variables. We produced two similar settings, with the only difference being the number of signal variables. The design matrix was standardised with mean 0 and variance 1, and the new continuous response variable was set as $\bm{y} = \bm{X} \bm{\beta} + \bm{\epsilon}$, where $\bm{\beta} = \{1,1,1,1,1,1,1,1,1,1,0,\dots,0\}$ for setting one, and $\bm{\beta} = \{1,1,1,1,1,0,\dots,0\}$ for setting two. In both settings, $\bm{\beta} = [\bm{\beta}_\mathcal{S}^\top,\bm{\beta}_\mathcal{N}^\top]^\top$, and the error term $\bm{\epsilon}$ was randomly generated such that $\bm{\epsilon} \sim N(\bm{0},\sigma^2\bm{I})$, where $\sigma^2$ were calculated to maintain an SNR of $1$ and $3$. For the first setting, $|\bm{\beta}_\mathcal{S}| = 10$ was selected to account for all $10$ baseline variables. We further decided to use  $|\bm{\beta}_\mathcal{S}| = 5$ as half of the total baseline variables. 

\begin{figure}[htbp]
    \centering
    \includegraphics[width = 0.9\textwidth]{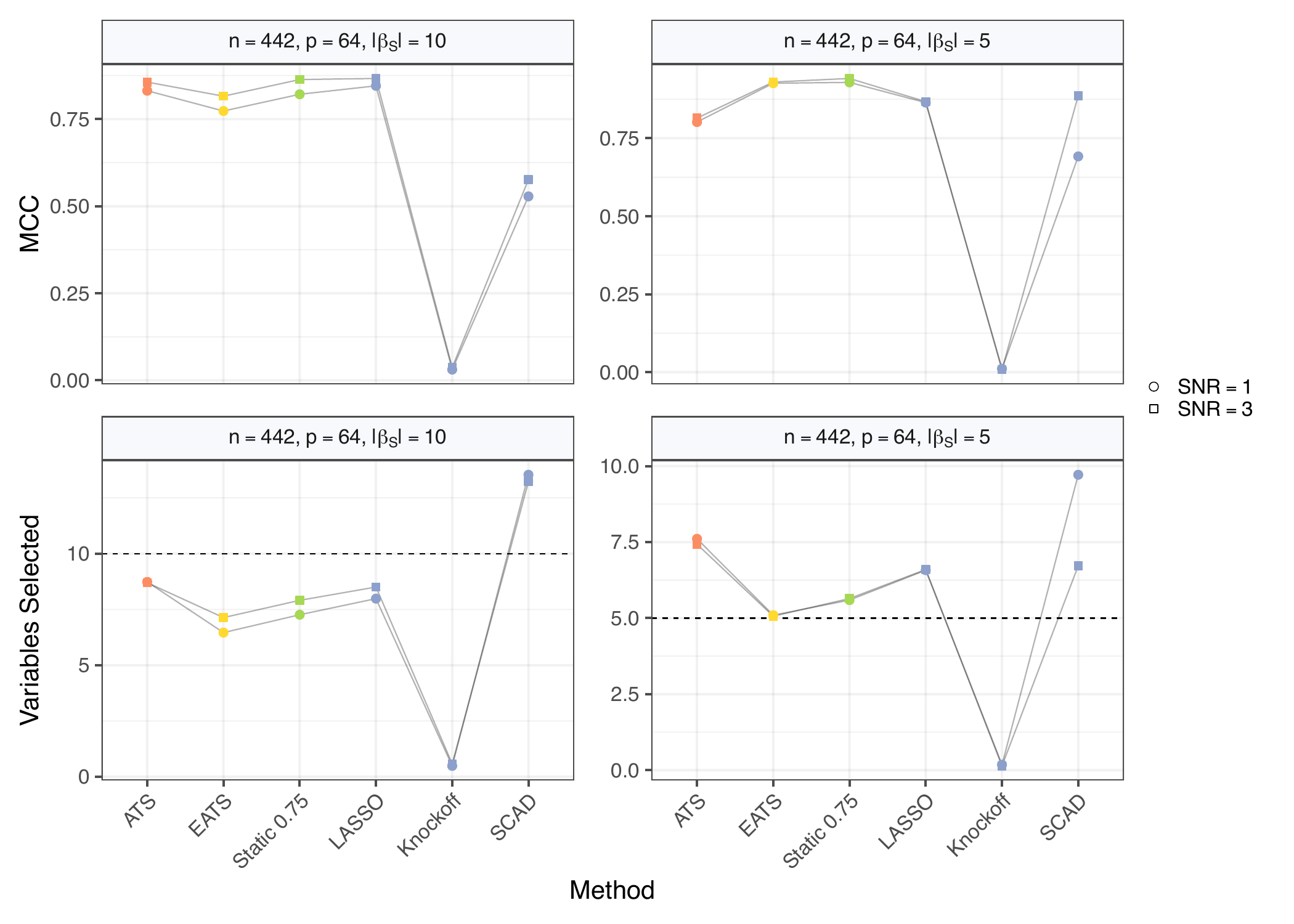} 
    \caption{Diabetes simulations ($n = 442, p = 64$) simulations for $|\bm{\beta}_{\mathcal{S}}| = 5$ and $10$ with SNR $=1 $ and $3$. The dotted line denotes the expected number of selected variables.}
    \label{fig:dia}
\end{figure}

Our empirical analysis in Section \ref{sec:simulation} shows that ATS has a higher average MCC than EATS for $p \leq n$. However, in the low-dimensional $p < n$ diabetes data, ATS only outperforms EATS in terms of MCC for the $|\bm{\beta}_{\mathcal{S}}|=10$ setting, as shown in Figure \ref{fig:dia}. In fact, for $|\bm{\beta}_{\mathcal{S}}|=5$, EATS performs considerably better than ATS, especially in selecting an appropriate number of variables.

\subsection{Error Control}

This section demonstrates that the ATS-derived $\pi$, combined with a user-defined $\E(V)$, can effectively satisfy the error control function. These parameters were applied to an artificially generated dataset with the same dimensions and coefficients outlined in Section \ref{sec:simulation}. 

When applying ATS to high-dimensional data, $\hat{\pi}(\hat{w})$ occasionally falls below $0.5$, failing to meet the $\pi$ requirements outlined in Section \ref{sec:errcontrol} and \citet{meinshausen_stability_2010}. To address this, we restricted the minimum value of $\hat{\pi}(\hat{w})$ to $0.501$, a limitation that primarily affects setting (I). Each simulation had a set $\E(V)$ and ATS-selected $\hat{\pi}(\hat{w})$. The results of this simulation study can be found in Table \ref{tab:err}. 

We provide three statistics to measure the performance of the false discovery rate properties. We define error bound satisfaction (EBS) as the proportion of cases out of $1000$ simulations where the number of falsely selected variables does not exceed the specified $\E(V)$. The true positive rate (TPR) is the ratio of correctly identified signal variables and the total number of selected variables. Lastly, we also give the average number of selected variables.

\begin{table}[!htbp]
\centering
\begin{tabular}{cccccc} 
\toprule
\textbf{Design Setting} &  \textbf{SNR} & \textbf{$\E(V)$} & \textbf{EBS} & \textbf{TPR} & \textbf{Avg. Selected Variables} \\
\midrule
\multirow{6}{*}{\shortstack{(I)\\ \\ $n = 20,\ p = 1000,$\\ $|\bm{\beta}_\mathcal{S}| = 2$}} 
                     & \multirow{3}{*}{1} & 2  & 1.00& 0.01 & 0.01 \\
                     &                    & 5 & 1.00 & 0.01 & 0.02\\
                     &                    & 10  & 1.00 & 0.01 & 0.03\\
                     \cmidrule(lr){2-6}
                     & \multirow{3}{*}{3}& 2 &1.00  & 0.06 & 0.12\\
                     &                    & 5 & 1.00 & 0.07 & 0.14\\
                     &                    & 10 &  1.00& 0.09 & 0.18\\
\midrule
\multirow{6}{*}{\shortstack{(II)\\ \\ $n = 100,\ p = 500,$\\ $|\bm{\beta}_\mathcal{S}| = 10$}}
                     & \multirow{3}{*}{1} & 2  & 1.00 & 0.17 & 1.79\\
                     &                    & 5 & 1.00 & 0.20 & 2.19\\
                     &                    & 10  &  1.00& 0.21 & 2.44\\
                     \cmidrule(lr){2-6}
                     & \multirow{3}{*}{3}& 2 & 1.00 & 0.33 & 3.36\\
                     &                    & 5 &  1.00& 0.36 & 3.94\\
                     &                    & 10 & 1.00 & 0.38 & 4.25\\
\midrule
\multirow{6}{*}{\shortstack{(III)\\ \\ $n = 200,\ p = 200,$\\ $|\bm{\beta}_\mathcal{S}| = 20$}}
                      & \multirow{3}{*}{1} & 2  & 1.00 & 0.18 & 3.68\\
                      &                    & 5 & 1.00 & 0.25 & 5.12\\
                      &                    & 10  & 1.00 & 0.29 & 6.26\\
                      \cmidrule(lr){2-6}
                      & \multirow{3}{*}{3}& 2 & 1.00 & 0.30 & 6.02\\
                      &                    & 5 & 1.00 & 0.37 & 7.59\\
                      &                    & 10 & 1.00 & 0.43 & 8.81\\
\midrule
\multirow{6}{*}{\shortstack{(IV)\\ \\ $n = 500,\ p = 100,$\\ $|\bm{\beta}_\mathcal{S}| = 20$}} 
                      & \multirow{3}{*}{1} & 2  & 1.00 & 0.30 & 5.98\\
                      &                    & 5 & 1.00 & 0.40 & 8.01\\
                      &                    & 10  & 1.00 & 0.50 & 10.10\\
                      \cmidrule(lr){2-6}
                      & \multirow{3}{*}{3}& 2 & 1.00 & 0.39 & 7.79\\
                      &                    & 5 & 1.00 & 0.59 & 11.83\\
                      &                    & 10 &  1.00& 0.70 & 14.12 \\
\bottomrule
\end{tabular}
\caption{Percentage of error control bound satisfaction, average true positive rate (TPR), and average number of selected variables across 1000 repetitions for the artificial datasets outlined in Section \ref{sec:simulation}. All simulations were conducted using EATS.}
\label{tab:err}
\end{table}

In order to restrain the number of falsely selected variables to be under the provided $\E(V)$, EATS is conservative in selecting variables. Consequently, all settings have an average number of selected variables much lower than the number of signal variables in $|\bm{\beta}_{\mathcal{S}}|$. In particular, in setting (I), the EATS-estimated $\pi$ is less than $0.5$ in 88\% of simulations and is therefore set to $0.501$ in order to satisfy the error bound. While the bound is always satisfied, a practitioner should weigh the benefits between underselection and error control. For example, setting (I) in Figure \ref{fig:nplothard} demonstrates EATS selecting a healthier range of variables, as error control is not enforced. Furthermore, the empirical distribution of $\pi$ for setting (I) given in Figure \ref{fig:pisims} is centred around $0.2$, which allows EATS to frequently select the correct number of variables. Constraining $\pi$ to $0.5$ leads to a large underselection of variables.

No combination of setting and SNR values in Table \ref{tab:err} produces a total number of falsely selected variables that exceeds the tolerated error $\E(V)$. Hence, we see that the error bound holds in our simulations. This analysis demonstrates that the error control properties in \citet{meinshausen_stability_2010} are not compromised when utilising EATS. Furthermore, while using EATS, the error control procedure is simplified, as only one parameter between $\E(V)$ and $q_\Lambda$ needs to be specified.

\section{Discussion and Conclusion}\label{chap:conclusion}
Stable threshold values chosen with accompanying data-driven techniques can considerably improve the results of stability selection. By implementing automatic elbow detection through a scree plot, we presented EATS, a method for estimating the stable threshold parameter $\pi$, which can adapt to various datasets and their respective variable selection probabilities. EATS is advantageous when selecting $\pi$, as it generates a set of variables that are uncorrelated with the outcome variable to estimate the selection probabilities of noise variables. 

Since EATS is required to apply stability selection to both the shuffled and original datasets, the computational time is approximately doubled, in comparison to regular stability selection. We believe that this is justifiable due to the promising results displayed in our simulations. Furthermore, the computation time for larger datasets is limited to a few minutes, while for smaller datasets, seconds.

EATS, however faces some limitations. Firstly, given the set-up of the automatic elbow detection method, the last candidate variable cannot be chosen as an elbow. If the last variable was considered as a candidate elbow, there would be an empty second group in $\bm{\delta}_2$. Secondly, the performance of EATS depends by construction on stability selection. EATS itself is not a variable selection algorithm, but is used to amplify and automate stability selection. If stability selection or the selected variable selection algorithm is unable to correctly identify signal variables, neither will EATS. As a last note, Section \ref{sec:errcontrol} demonstrated that EATS does not effect the error control property of stability selection, as long as the estimated threshold is greater than $0.5$. One strength of EATS is that it has the capability of detecting signal variables with small selection probabilities, less than $0.5$. Since this is the case, EATS may correctly select the variables, however surrender the theoretical error bound. While we do not show this example, EATS can also be applied to other stability selection extensions, such as Complementary Pairs Stability Selection, where the error can be controlled with threshold values less than $0.5$ \citep{shah_variable_2013}.

In this paper, we compared the results of ATS and EATS, against common variable selection algorithms and a static value of $\pi$. Through an extensive simulation study including artificially generated data and real world datasets, we found that both ATS and EATS perform similarly to or better than the midpoint of the recommended range of $\pi$ values. Additionally, ATS and EATS eliminate the need for users to specify the stable threshold parameter. Our analysis also found that EATS thrives in high dimensions when the number of predictor variables is greater than the sample size. When the number of predictor variables is smaller than the sample size, EATS and ATS are comparable. Therefore, we recommend using EATS as a default method. Moreover, we also demonstrated that the error control bound remains intact when deploying our methods, whilst providing guidance on how to utilise such properties.

As the performance of stability selection is heavily dependent on the stable threshold parameter $\pi$, it is important to consider and analyse the variables' selection probabilities, to estimate $\pi$ accordingly. To overcome this challenge, we recommend that EATS be the default method to aid a practitioner when utilising stability selection. Still, it is important to remember that the performance of EATS is dependent on the performance of stability selection. 

Since EATS does not require tuning and can adapt to selection probabilities from a wide range of dimensions, it present itself as an easy and useful supplement to estimating the stable threshold parameter within the stability selection framework.

\section*{Acknowledgements}

Samuel Muller and Garth Tarr were supported by the Australian Research Council (DP210100521). Garth Tarr was supported in part by the AIR@innoHK program of the Innovation and Technology Commission of Hong Kong. The authors would like to thank the Associate Editor and the two reviewers for their helpful comments.

\clearpage

\bibliographystyle{apalike} 

\begin{thebibliography}{}

\bibitem[Alexander and Lange, 2011]{alexander_stability_2011}
Alexander, D.~H. and Lange, K. (2011).
\newblock Stability selection for genome-wide association.
\newblock {\em Genetic Epidemiology}, 35(7):722--728.

\bibitem[Barber and Candès, 2015]{barber_controlling_2015}
Barber, R.~F. and Candès, E.~J. (2015).
\newblock Controlling the false discovery rate via knockoffs.
\newblock {\em The Annals of Statistics}, 43(5):2055--2085.

\bibitem[Bodinier et~al., 2023]{bodinier_automated_2023}
Bodinier, B., Filippi, S., Nøst, T.~H., Chiquet, J., and Chadeau-Hyam, M. (2023).
\newblock Automated calibration for stability selection in penalised regression and graphical models.
\newblock {\em Journal of the Royal Statistical Society Series C: Applied Statistics}, 72(5), 1375–1393

\bibitem[Breheny and Huang, 2011]{breheny_coordinate_2011}
Breheny, P. and Huang, J. (2011).
\newblock Coordinate descent algorithms for nonconvex penalized regression, with applications to biological feature selection.
\newblock {\em Annals of Applied Statistics}, 5(1):232--253.

\bibitem[Bühlmann et~al., 2014]{buhlmann_high-dimensional_2014}
Bühlmann, P., Kalisch, M., and Meier, L. (2014).
\newblock High-dimensional statistics with a view toward applications in biology.
\newblock {\em Annual Review of Statistics and Its Application}, 1(1):255--278.

\bibitem[Candès et~al., 2018]{candes_panning_2018}
Candès, E., Fan, Y., Janson, L., and Lv, J. (2018).
\newblock Panning for gold: ‘model-X’ knockoffs for high dimensional controlled variable selection.
\newblock {\em Journal of the Royal Statistical Society Series B: Statistical Methodology}, 80(3):551--577.

\bibitem[Cattell, 1966]{cattell_scree_1966}
Cattell, R.~B. (1966).
\newblock The scree test for the number of factors.
\newblock {\em Multivariate Behavioral Research}, 1(2):245--276.


\bibitem[Efron et~al., 2004]{efron_least_2004}
Efron, B., Hastie, T., Johnstone, I., and Tibshirani, R. (2004).
\newblock Least angle regression.
\newblock {\em The Annals of Statistics}, 32(2):407--451.

\bibitem[Efron, 2022]{efron_lars_2022}
Efron, T. H. a.~B. (2022).
\newblock lars: {Least} {Angle} {Regression}, {Lasso} and {Forward} {Stagewise}.

\bibitem[Fan and Li, 2001]{fan_variable_2001}
Fan, J. and Li, R. (2001).
\newblock Variable selection via nonconcave penalized likelihood and its oracle properties.
\newblock {\em Journal of the American Statistical Association}, 96(456):1348--1360.


\bibitem[Fix and Hodges, 1989]{fix_discriminatory_1989}
Fix, E. and Hodges, J.~L. (1989).
\newblock Discriminatory analysis. Nonparametric discrimination: consistency properties.
\newblock {\em International Statistical Review / Revue Internationale de Statistique}, 57(3):238--247.

\bibitem[Friedman et~al., 2010]{friedman_regularization_2010}
Friedman, J., Tibshirani, R., and Hastie, T. (2010).
\newblock Regularization paths for generalized linear models via coordinate descent.
\newblock {\em Journal of Statistical Software}, 33(1):1--22.

\bibitem[Gauraha et~al., 2017]{gauraha_post_2017}
Gauraha, N., Pavlenko, T., and Parui, S.~k. (2017).
\newblock Post lasso stability selection for high dimensional linear models.
\newblock {\em Proceedings of the 6th {International} {Conference} on {Pattern} {Recognition} {Applications} and {Methods}}, pages 638--646, Porto, Portugal.

\bibitem[Hofner and Hothorn, 2021]{hofner_stabs_2021}
Hofner, B. and Hothorn, T. (2021).
\newblock stabs: stabilitys selection with error control.

\bibitem[Hédou et~al., 2024]{hedou_discovery_2024}
Hédou, J., Marić, I., Bellan, G., Einhaus, J., Gaudillière, D.~K., Ladant, F.-X., Verdonk, F., Stelzer, I.~A., Feyaerts, D., Tsai, A.~S., Ganio, E.~A., Sabayev, M., Gillard, J., Amar, J., Cambriel, A., Oskotsky, T.~T., Roldan, A., Golob, J.~L., Sirota, M., Bonham, T.~A., Sato, M., Diop, M., Durand, X., Angst, M.~S., Stevenson, D.~K., Aghaeepour, N., Montanari, A., and Gaudillière, B. (2024).
\newblock Discovery of sparse, reliable omic biomarkers with {Stabl}.
\newblock {\em Nature Biotechnology}, 42(10):1581--1593.

\bibitem[Lu et~al., 2017]{lu_performance_2017}
Lu, D., Weljie, A., de~Leon, A.~R., McConnell, Y., Bathe, O.~F., and Kopciuk, K. (2017).
\newblock Performance of variable selection methods using stability-based selection.
\newblock {\em BMC Research Notes}, 10(1):143.

\bibitem[MacQueen, 1967]{macqueen_methods_1967}
MacQueen, J. (1967).
\newblock Some methods for classification and analysis of multivariate observations.
\newblock {\em Proceedings of the {Fifth} {Berkeley} {Symposium} on {Mathematical} {Statistics} and {Probability}, {Volume} 1: {Statistics}}, 5.1,281--298

\bibitem[Maddu et~al., 2022]{maddu_stability_2022}
Maddu, S., Cheeseman, B.~L., Sbalzarini, I.~F., and Müller, C.~L. (2022).
\newblock Stability selection enables robust learning of differential equations from limited noisy data.
\newblock {\em Proceedings of the Royal Society A: Mathematical, Physical and Engineering Sciences}, 478(2262):20210916.

\bibitem[Matthews, 1975]{matthews_comparison_1975}
Matthews, B.~W. (1975).
\newblock Comparison of the predicted and observed secondary structure of {T4} phage lysozyme.
\newblock {\em Biochimica et Biophysica Acta (BBA) - Protein Structure}, 405(2):442--451.


\bibitem[Meinshausen and Bühlmann, 2010]{meinshausen_stability_2010}
Meinshausen, N. and Bühlmann, P. (2010).
\newblock Stability selection.
\newblock {\em Journal of the Royal Statistical Society Series B: Statistical Methodology}, 72(4):417--473.

\bibitem[Melikechi and Miller, 2024]{melikechi_2024}
Melikechi, O. and Miller, J. W. (2024).
\newblock Integrated path stability selection.
\newblock {\em  arXiv. https://doi.org/10.48550/arXiv.2403.15877.}


\bibitem[Mordelet et~al., 2013]{mordelet_stability_2013}
Mordelet, F., Horton, J., Hartemink, A.~J., Engelhardt, B.~E., and Gordân, R. (2013).
\newblock Stability selection for regression-based models of transcription factor–{DNA} binding specificity.
\newblock {\em Bioinformatics}, 29(13):i117--i125.

\bibitem[Patterson and Sesia, 2022]{patterson_knockoff_2022}
Patterson, E. and Sesia, M. (2022).
\newblock knockoff: {The} knockoff filter for controlled variable selection.

\bibitem[{R Core Team}, 2024]{r_core_team_r_2024}
{R Core Team} (2024).
\newblock {\em R: {A} language and environment for statistical computing}.
\newblock R Foundation for Statistical Computing, Vienna, Austria.

\bibitem[Rumer et~al., 2022]{rumer_integrated_2022}
Rumer, K.~K., Hedou, J., Tsai, A., Einhaus, J., Verdonk, F., Stanley, N., Choisy, B., Ganio, E., Bonham, A., Jacobsen, D., Warrington, B., Gao, X., Tingle, M., McAllister, T.~N., Fallahzadeh, R., Feyaerts, D., Stelzer, I., Gaudilliere, D., Ando, K., Shelton, A., Morris, A., Kebebew, E., Aghaeepour, N., Kin, C., Angst, M.~S., and Gaudilliere, B. (2022).
\newblock Integrated single-cell and plasma proteomic modeling to predict surgical site complications, a prospective cohort study.
\newblock {\em Annals of surgery}, 275(3):582.

\bibitem[Shah and Samworth, 2013]{shah_variable_2013}
Shah, R.~D. and Samworth, R.~J. (2013).
\newblock Variable selection with error control: another look at stability selection.
\newblock {\em Journal of the Royal Statistical Society. Series B (Statistical Methodology)}, 75(1):55--80.

\bibitem[Shi et~al., 2021]{shi_quantitative_2021}
Shi, C., Wei, B., Wei, S., Wang, W., Liu, H., and Liu, J. (2021).
\newblock A quantitative discriminant method of elbow point for the optimal number of clusters in clustering algorithm.
\newblock {\em EURASIP Journal on Wireless Communications and Networking}, 2021(1):31.

\bibitem[Thorndike, 1953]{thorndike_who_1953}
Thorndike, R.~L. (1953).
\newblock Who belongs in the family?
\newblock {\em Psychometrika}, 18:267--276.

\bibitem[Tibshirani, 1996]{tibshirani_regression_1996}
Tibshirani, R. (1996).
\newblock Regression shrinkage and selection via the lasso.
\newblock {\em Journal of the Royal Statistical Society. Series B (Methodological)}, 58(1):267--288.

\bibitem[Yuan and Yang, 2019]{yuan_research_2019}
Yuan, C. and Yang, H. (2019).
\newblock Research on {K}-value selection method of {K}-means clustering algorithm.
\newblock {\em J}, 2(2):226--235.

\bibitem[Zhao and Yu, 2006]{zhao_lasso_2006}
Zhao, P. and Yu, B. (2006).
\newblock On Model Selection Consistency of Lasso.
\newblock {\em Journal of Machine Learning Research}, 7(90), 2541–2563.



\bibitem[Zhu and Ghodsi, 2006]{zhu_automatic_2006}
Zhu, M. and Ghodsi, A. (2006).
\newblock Automatic dimensionality selection from the scree plot via the use of profile likelihood.
\newblock {\em Computational Statistics \& Data Analysis}, 51(2):918--930.



\end{thebibliography}

\clearpage
\appendix

\section{Table of Definitions}\label{app:defs}
\begin{table}[H]
\centering
\renewcommand{\arraystretch}{1.2}
\begin{tabularx}{\textwidth}{lX}
\toprule
\textbf{Symbol} & \textbf{Description} \\
\midrule
$\bm{X} \in \mathbb{R}^{n \times p}$ & Design matrix with $n$ samples and $p$ predictors \\
$\bm{y} \in \mathbb{R}^{n}$ & Response vector \\
$\bm{\beta}^\top = (\beta_1, \dots, \beta_p)$ & True regression coefficient vector \\
$\bm{\epsilon}$ & Random noise vector \\
$\mathcal{S} = \{k: \beta_k \neq 0,\ k =1,\dots,p\}$ & Index set of signal (non-zero) variables \\
$\mathcal{N} = \{k :\beta_k = 0,\ k = 1,\dots,p\}$ & Index set of noise (zero) variables \\
$\bm{\beta}_{\mathcal{S}}, \bm{\beta}_{\mathcal{N}}$ & Coefficient for signal and noise variables respectively \\
$\bm{X}_{\mathcal{S}}, \bm{X}_{\mathcal{N}}$ & Matrices of $\bm{X}$ containing signal and noise predictors respectively \\
$n, p$ & Sample size and number of predictors \\
$\mathbbm{1}$ & An indicator function \\
$\bm{\Sigma} \in \mathbb{R}^{p\times p}$ & A covariance matrix \\
$|\cdot|$ & Cardinality of a set \\
$\text{SNR} = \lVert \bm{X} \bm{\beta}\rVert ^2 _2 /n\sigma^2$ & Signal-to-noise ratio \\
\bottomrule
\end{tabularx}
\caption{General data and model setup notation.}
\end{table}

\begin{table}[H]
\centering
\renewcommand{\arraystretch}{1.2}
\begin{tabularx}{\textwidth}{lX}
\toprule
\textbf{Symbol} & \textbf{Description} \\
\midrule
$\lambda$ & Regularisation parameter \\
$\Lambda =  \{\lambda_1, \ldots, \lambda_t\}$ & Set of regularisation parameters \\
$\hat{\bm{\beta}}^{\lambda}$ & Estimated coefficient vector for given $\lambda$ \\
$\hat{\mathcal{S}}^{\lambda}$ & Set of variables selected at penalty $\lambda$ \\
$\hat{\mathcal{S}}^\Lambda = \bigcup_{\lambda \in \Lambda} \hat{\mathcal{S}}^\lambda$ & Union of selected variable sets across $\Lambda$ \\
$\mathcal{I}$ & Random subsample index set \\
$B$ & Number of subsamples in stability selection \\
$q_\Lambda = \E(|\hat{\mathcal{S}}^\Lambda (\mathcal{I})|)$ & Average number of selected variables across all $\lambda \in \Lambda$ \\
\bottomrule
\end{tabularx}
\caption{Variable selection notation.}
\end{table}

\begin{table}[H]
\centering
\renewcommand{\arraystretch}{1.2}
\begin{tabularx}{\textwidth}{lX}
\toprule
\textbf{Symbol} & \textbf{Description} \\
\midrule
$\max_{\lambda \in \Lambda} (\hat{\Pi}^\lambda _k )$ & Maximum empirical variable selection probability \\
$\pi$ & Stable threshold parameter \\
$\hat{\mathcal{S}}^{\text{stable}} = \{k :\max_{\lambda \in \Lambda} (\hat{\Pi}^\lambda _k )\geq \pi; k=1,\ldots,p\}$ & Set of stable and selected variables \\
$V = |\mathcal{N} \cap \hat{\mathcal{S}}^{\text{stable}} |$ & Number of falsely selected variables \\
$\E(V)$ & Expected number of falsely selected variables \\
\bottomrule
\end{tabularx}
\caption{Stability Selection notation.}
\end{table}

\begin{table}[H]
\centering
\renewcommand{\arraystretch}{1.2}
\begin{tabularx}{\textwidth}{lX}
\toprule
\textbf{Symbol} & \textbf{Description} \\
\midrule
$\hat{d}_{\sigma(j)} = \max_{\lambda \in \Lambda} (\hat{\Pi}^\lambda _{\sigma(j)} ) $ & Ordered empirical selection probability \\
$\Delta = \{\hat{d}_{1}, \hat{d}_2,\dots, \hat{d}_p\}$ & Non-increasing vector of selection probabilities \\
$w$ & Candidate elbow index \\
$\Omega = \{l(1),\dots,l(p-1)\},\ w = 1,\dots,p-1$ & Set of profile log-likelihoods for candidate elbows \\
$l(w)$ & Profile log-likelihood at elbow $w$ \\
$\hat{w} = \argmax_w \Omega$ & Estimated elbow index \\
$\hat{\pi}(\hat{w}) = \Delta[\hat{w}]$ & EATS-estimated stable threshold \\
$(\bm{X}^*, \bm{y}^*)$ & Shuffled dataset used to estimate exclusion probabilities \\
$\eta = \hat{F}^{-1} _{\Delta^*} (0.95).$ & Exclusion probability threshold \\
$\tilde{p}$ = $\max \{j\ |\ \hat{d}_j \geq \eta,\ j = 1,\dots, p\}$ & Number of variables above exclusion threshold \\
$\tilde{\Delta} = \{\hat{d}_{1}, \hat{d}_2,\dots, \hat{d}_{\tilde{p}}\}$ & Filtered selection probabilities \\
\bottomrule
\end{tabularx}
\caption{ATS and EATS algorithm notation.}
\end{table}

\section{Additional Derivations}\label{app:theorems}
\newpage
Here we present some additional information regarding the ATS algorithm from Section \ref{sec:ats} and proof of Equation \ref{eq:MLE}.

To solve for $\hat{w}$ we require a pre-determined distribution $f$, where a normal distribution is recommended in \citet{zhu_automatic_2006}. Therefore, assume $\bm{\delta}_1\text{ and }\bm{\delta}_2 \sim \mathcal{N}(\mu, \sigma^2)$, then $\bm{\theta}_{v} = \{\mu_v, \sigma_v\},\ v =1,2$ for $\delta_v$. By extending this notation, $\hat{\bm{\theta}}_v = \{\hat{\mu}_v, \hat{\sigma}_v\},\ v = 1,2$. Assuming such a distribution, the profile log-likelihood function for the entire set of variables is given as 
\begin{align*}
    l (w) &=  \sum^w _{i=1} \log f(\hat{d}_i; \hat{\bm{\theta}}_1(w)) + \sum^p _{m=w+1} \log f(\hat{d}_m ; \hat{\bm{\theta}}_2(w)) \nonumber\\ 
                      &= \log\left((\hat{\sigma}\sqrt{2\pi})^{-w}\right) - \frac{1}{2\hat{\sigma}^2}\sum^w_{i=1} (\hat{d}_i - \hat{\mu}_1)^2 
                       + \\ 
                       & \qquad \qquad \log\left((\hat{\sigma}\sqrt{2\pi})^{-p}\right) - \frac{1}{2\hat{\sigma}^2}\sum^p _{m=w + 1} (\hat{d}_m - \hat{\mu}_2)^2 \nonumber\\
                      &= (-p-w)\log(\hat{\sigma} \sqrt{2\pi}) - \frac{1}{2\hat{\sigma}^2} \left(\sum^w _{i=1} (\hat{d}_i - \hat{\mu}_1 )^2 + \sum^p _{m=w+1} (\hat{d}_m - \hat{\mu}_2)^2 \right).
\end{align*}
Note here that $\pi$ refers to the $\pi$ given in the normal distribution function, not the stable threshold parameter. Choosing to use the subset of candidate elbows in EATS, we simply replace $p$ with $\tilde{p}$. 

Here, $\hat{\mu}_1,\hat{\mu}_2$ are the MLE for $\mu_1$ and $\mu_2$ (sample averages) and $\hat{\sigma}^2$ is the MLE for the common scale parameter $\sigma^2$ (pooled estimate). For completion, we give the definitions for the pooled estimator
\begin{align}\label{eq:pooledvar}
    \hat{\sigma}^2 = \frac{\sum_{i= 1} ^w (\hat{d}_i - \hat{\mu}_1 )^2 + \sum_{m = w + 1} ^ p (\hat{d}_m - \hat{\mu}_2)^2}{p-2}.
\end{align}
Then, using the definition in Equation \ref{eq:pooledvar} allows us to simplify $l(w)$ as
$l (w) \propto (-p-w)\log (\hat{\sigma}\sqrt{2\pi})$, or $l (w) \propto (-p_{\eta}-w)\log (\hat{\sigma}\sqrt{2\pi})$. 

We did not explore the use of other distributions for $\bm{\delta}_1$ and $\bm{\delta}_2$ and is a potential avenue for further work. While the final equation is easy to implement via hand, any programming language can solve the first line  $\sum^w _{i=1} \log f(\hat{d}_i; \hat{\bm{\theta}}_1(w)) + \sum^p _{m=w+1} \log f(\hat{d}_m ; \hat{\bm{\theta}}_2(w))$ directly, via some density function (such as \textit{dnorm} in R), after the parameters in $\hat{\bm{\theta}}_v$ is computed.

\section{Simulation Study Additional Results}\label{app:sims}
\newpage
Here we specify some additional details and results from the simulation studies conducted in Section \ref{chap:numerical}.

The coefficients for the artificial datasets in Section \ref{sec:simulation} are given in Table \ref{tab:coefficients}.

\begin{table}[H] 
    \centering
    \begin{tabular}{ll}
        \toprule
        \textbf{Setting} & \textbf{Active Coefficients} \\
        \midrule
        (I) & 1, 1 \\
        (II) & -3, 2, -2, -3, -3, -1, -3, 2, 1, -2 \\
        (III) & -2, 1, -2, 2, -2, 2, 1, 2, -2, -2, -2, 1, -2, 2, -1, 2, 1, 1, -2, -2 \\
        (IV) & 2, -1, -2, 2, -1, -2, -2, -2, -1, 1, 2, -1, -2, -1, -3, 2, 1, -2, -2, 2 \\
        (Uniform) & -3, 3, 3, -3, 2, 1, -3, 1, 3, -3\\
        ($t_3$) & -2, -1, 1, -2, 3, -1, 3, -1, -1, -1\\
        \bottomrule
    \end{tabular}
    \caption{Coefficients of signal variables in the artificial datasets.}
    \label{tab:coefficients}
\end{table}

\begin{table}[H] 
    \centering
    \begin{tabular}{ll}
        \toprule
        \textbf{Setting} & \textbf{Active Coefficients} \\
        \midrule
        \shortstack{$p = \{50,100,500,1000\},$\\ $n = 100,\  |\bm{\beta}_{\mathcal{S}}| = 10$} & -2, -2, 1, -2, -2, -3, -2, -2, 3, -2\\
        
        \\
        
        \shortstack{$n = \{100,200,400,800\},$\\ $p = 200,\  |\bm{\beta}_{\mathcal{S}}| = 20$} & -1, -3, 2, 3, -3, -1, 2, 2, 2, -1, -1, 1, -2, 1, 1, 1, -1, -3, -3, 1\\
        \bottomrule
    \end{tabular}
    \caption{Coefficients of signal variables when varying $n$ and $p$.}
    \label{tab:coefficientsNP}
\end{table}


\begin{figure}[htbp]
    \centering
    \includegraphics[width = 0.9\textwidth]{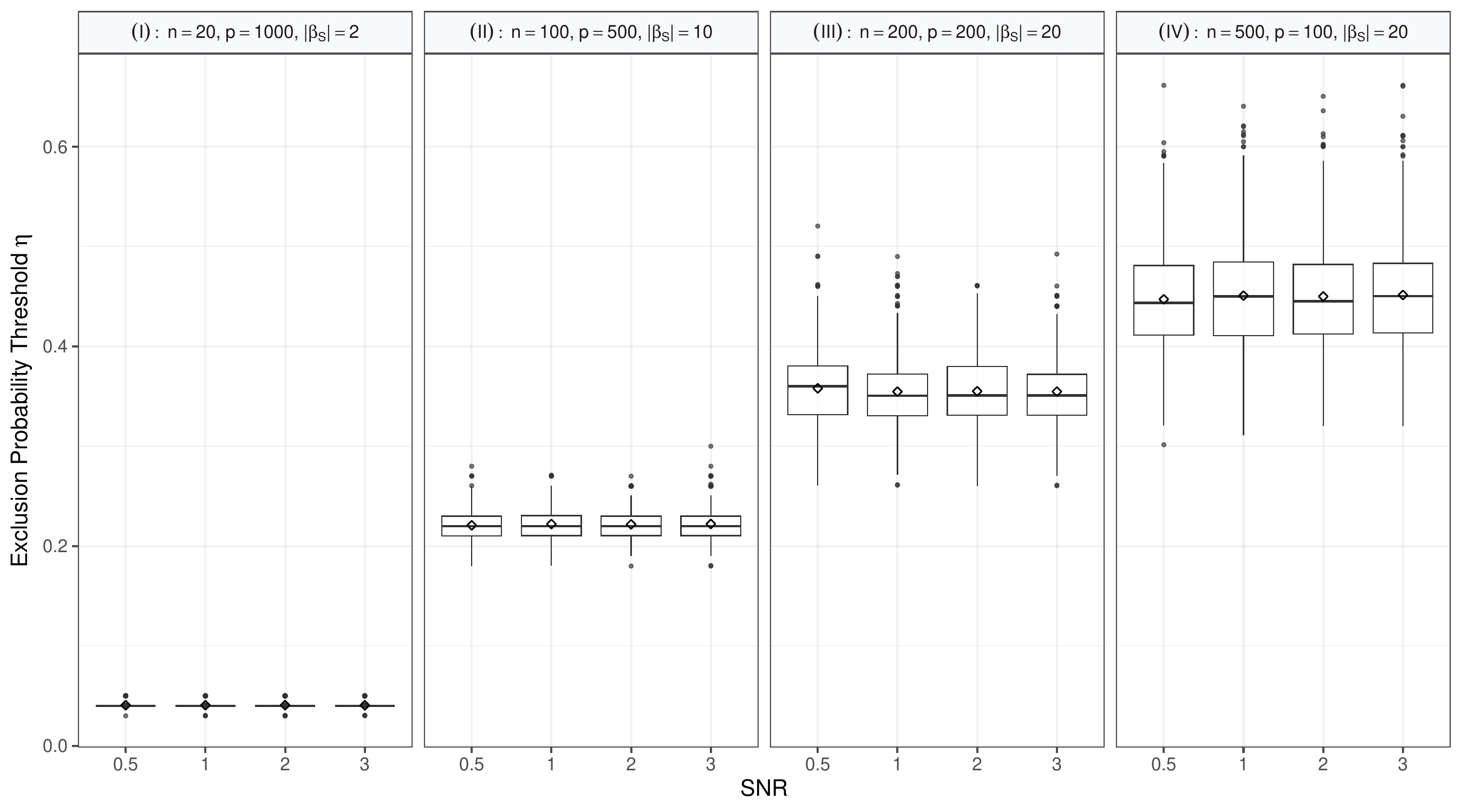} 
    \caption{Distribution of exclusion probability threshold $\eta$ for varying simulation studies (I) - (IV).}
    \label{fig:etasims}
\end{figure}

\begin{figure}[htbp]
    \centering
    \includegraphics[width = 0.9\textwidth]{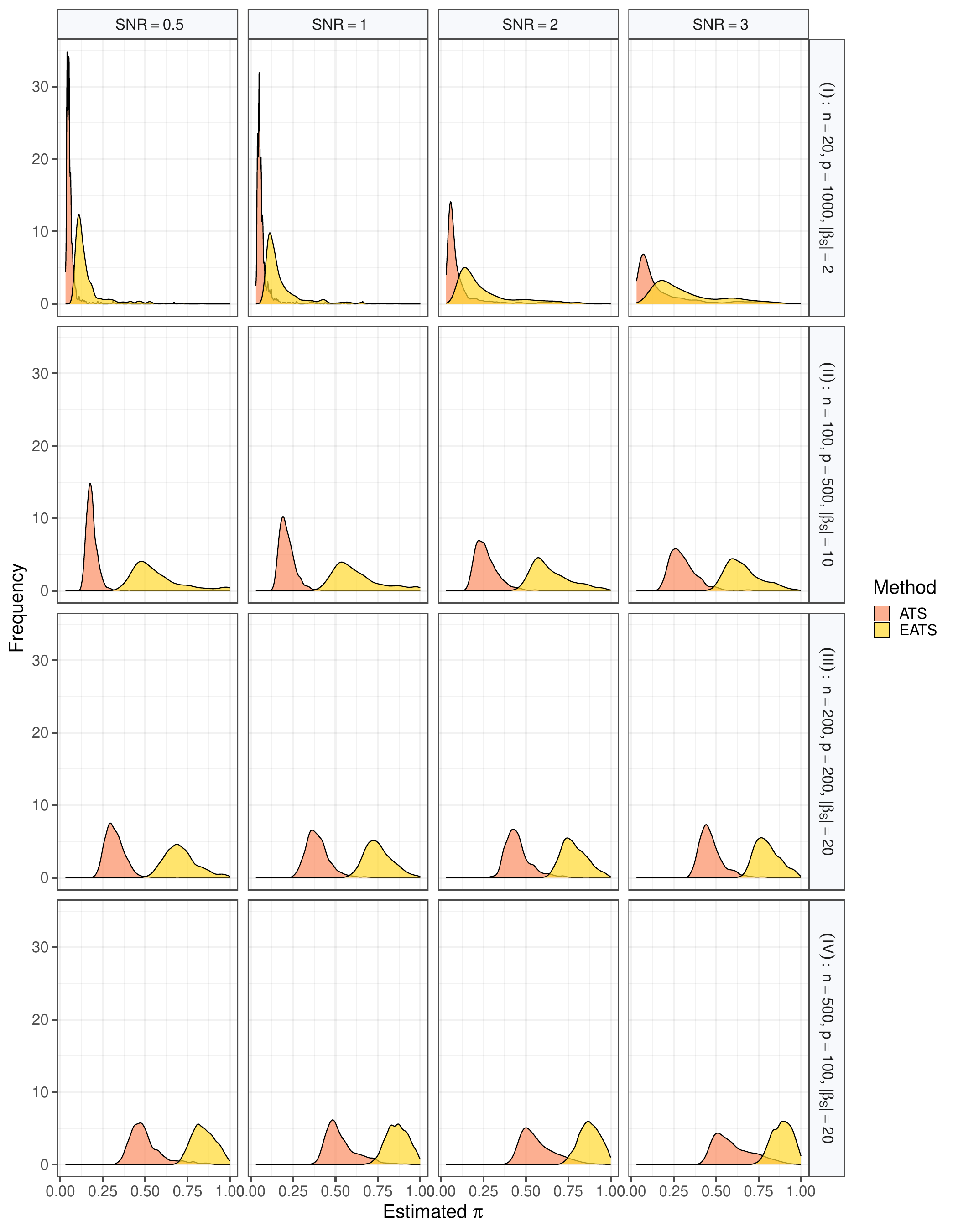} 
    \caption{Distribution of ATS and EATS selected $\hat{\pi}(\hat{w})$ threshold for varying simulation studies (I) - (IV).}
    \label{fig:pisims}
\end{figure}

\begin{figure}[htbp]
    \centering
    \includegraphics[width = 0.9\textwidth]{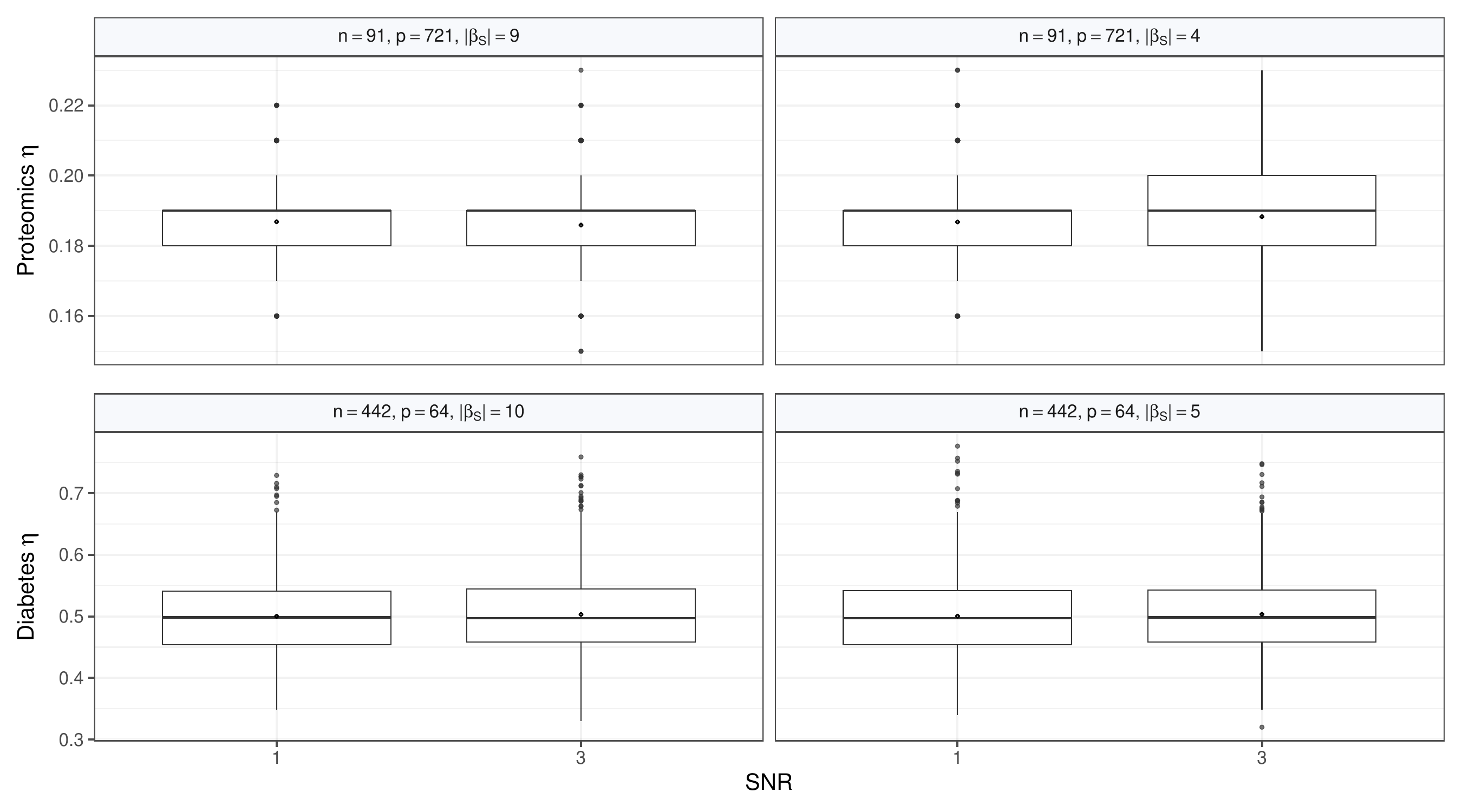} 
    \caption{Distribution of exclusion probability threshold $\eta$ for plasma proteomics and diabetes simulation studies.}
    \label{fig:exclprodia}
\end{figure}

\begin{figure}[htbp]
    \centering
    \includegraphics[width = 0.9\textwidth]{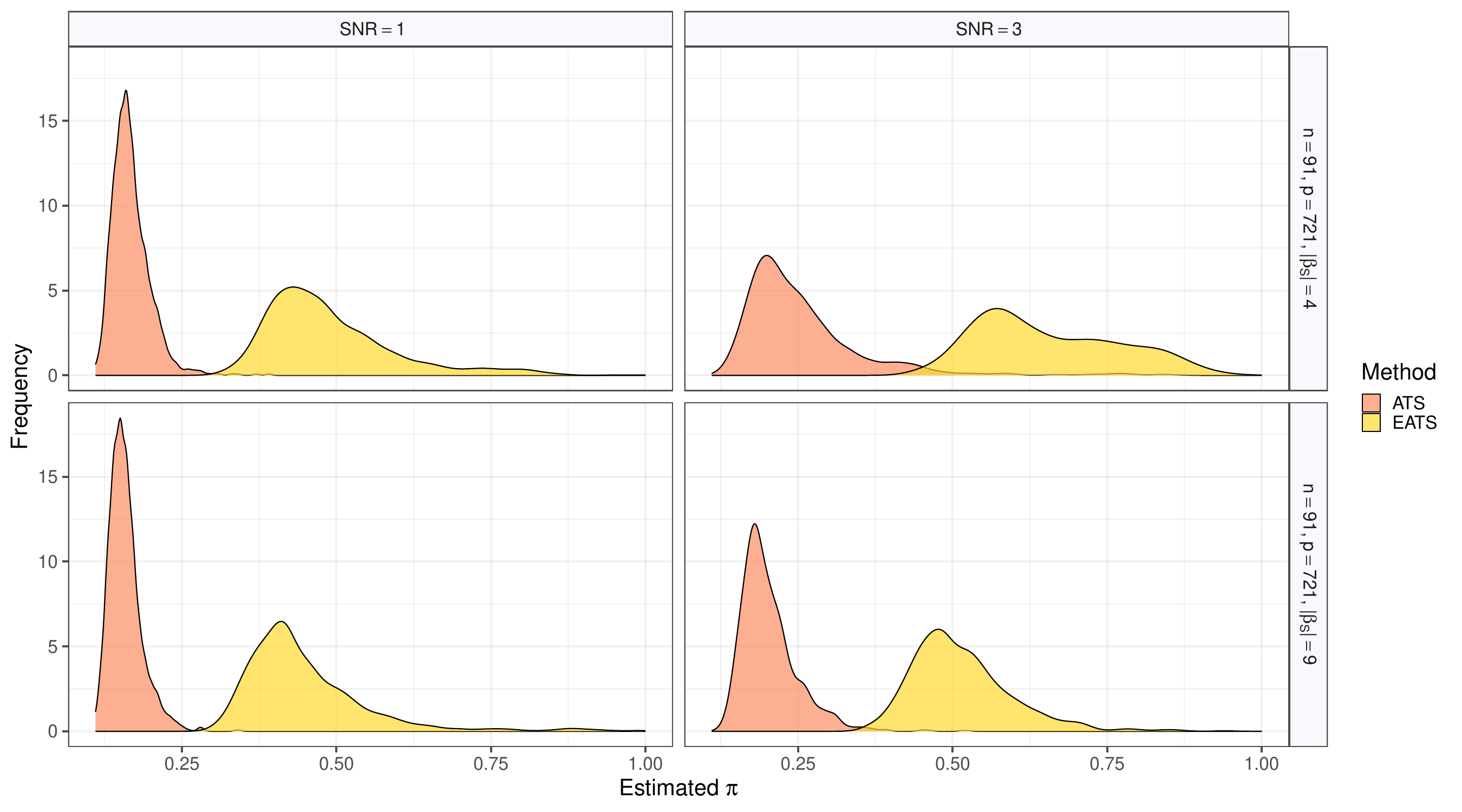} 
    \caption{Distribution of ATS and EATS selected $\hat{\pi}(\hat{w})$ threshold for proteomics dataset.}
    \label{fig:proPi}
\end{figure}

\begin{figure}[htbp]
    \centering
    \includegraphics[width = 0.9\textwidth]{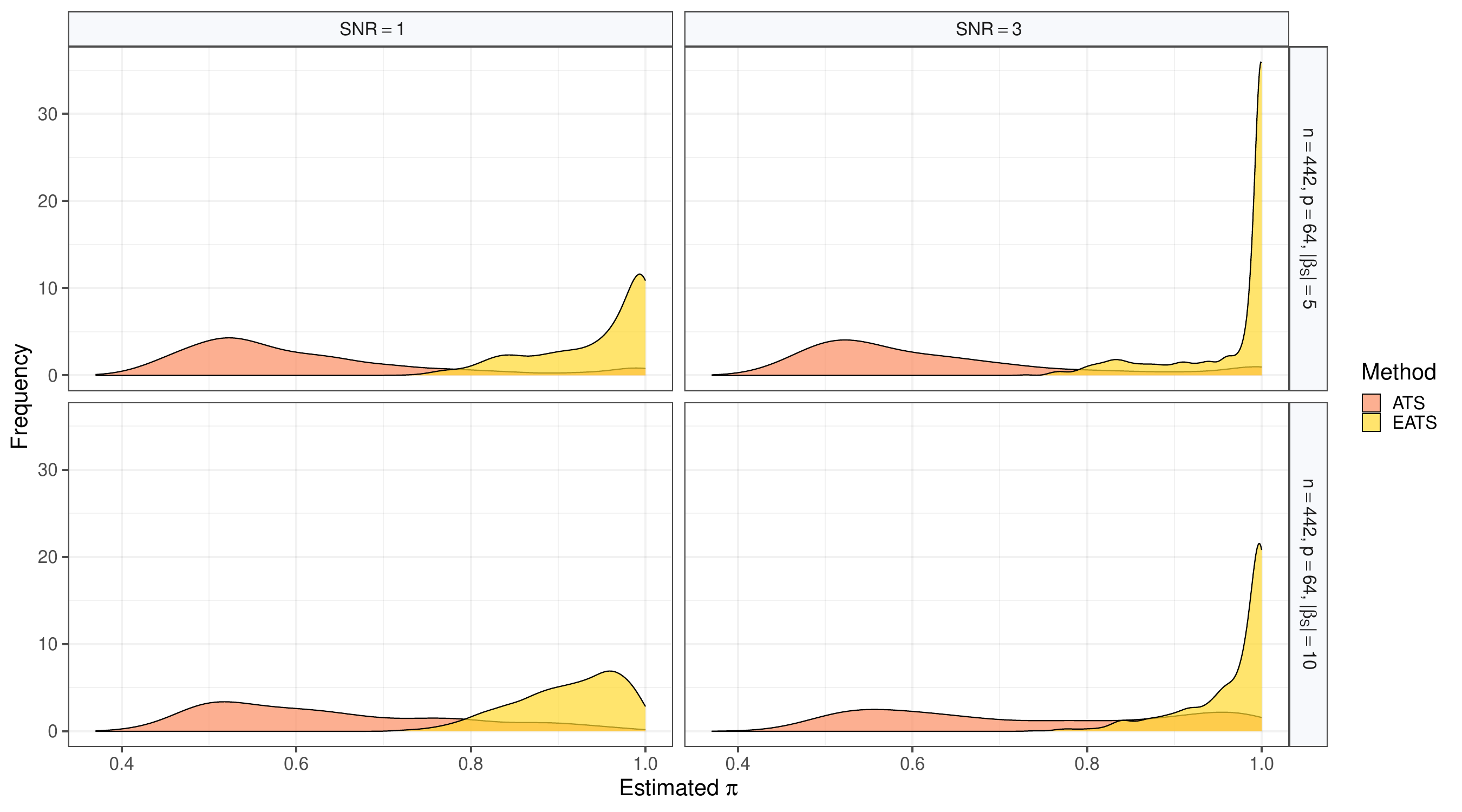} 
    \caption{Distribution of ATS and EATS selected $\hat{\pi}(\hat{w})$ threshold for diabetes dataset.}
    \label{fig:diabPi}
\end{figure}

\end{document}